\documentclass[aps,twocolumn,prb,floatfix,showpacs,lengthcheck,citeautoscript]{revtex4-1}
\usepackage{amsmath}
\usepackage{amssymb}
\usepackage{graphicx}
\usepackage{bm}
\usepackage{color}
\usepackage{times}
\usepackage{MnSymbol}
\usepackage{verbatim}
\usepackage{enumerate}

\newcommand{\bea}{\begin{eqnarray}}
\newcommand{\eea}{\end{eqnarray}}
\newcommand{\be}{\begin{equation}}
\newcommand{\ee}{\end{equation}}

\newcommand{\ci}{\mathrm{i}}

\newcommand{\ket}[1]{| #1 \rangle}
\newcommand{\Ket}[1]{| #1 \rangle\rangle}

\newcommand{\bra}[1]{\langle #1 |}

\begin{document}
\title{Time-resolved Hall conductivity of pulse-driven topological quantum systems}
\author{Lucila Peralta Gavensky}
\affiliation{Centro At{\'{o}}mico Bariloche and Instituto Balseiro,
Comisi\'on Nacional de Energ\'{\i}a At\'omica, 8400 Bariloche, Argentina}
\affiliation{Consejo Nacional de Investigaciones Cient\'{\i}ficas y T\'ecnicas (CONICET), Argentina}

\author{Gonzalo Usaj}
\affiliation{Centro At{\'{o}}mico Bariloche and Instituto Balseiro,
Comisi\'on Nacional de Energ\'{\i}a At\'omica, 8400 Bariloche, Argentina}
\affiliation{Consejo Nacional de Investigaciones Cient\'{\i}ficas y T\'ecnicas (CONICET), Argentina}

\author{C. A. Balseiro}
\affiliation{Centro At{\'{o}}mico Bariloche and Instituto Balseiro,
Comisi\'on Nacional de Energ\'{\i}a At\'omica, 8400 Bariloche, Argentina}
\affiliation{Consejo Nacional de Investigaciones Cient\'{\i}ficas y T\'ecnicas (CONICET), Argentina}

\begin{abstract}
We address the question of how the time-resolved bulk Hall response of a two dimensional honeycomb lattice develops when driving the system with a pulsed perturbation. A simple toy model that switches a valley Hall signal by breaking inversion symmetry is studied in detail for slow quasi-adiabatic ramps and sudden quenches, obtaining an oscillating dynamical response that depends strongly on doping and time-averaged values that are determined both by the out of equilibrium occupations and the Berry curvature of the final states. On the other hand, the effect of irradiating the sample with a circularly-polarized infrared pump pulse that breaks time reversal symmetry and thus ramps the system into a non-trivial topological regime is probed. Even though there is a non quantized average signal due to the break down of the Floquet adiabatical picture, some features of the photon-dressed topological bands are revealed to be present even in a few femtosecond timescale. Small frequency oscillations during the transient response evidence the emergence of dynamical Floquet gaps which are consistent with the instantaneous amplitude of the pump envelope. On the other hand, a characteristic heterodyining effect is manifested in the model. The presence of a remnant Hall response for ultra-short pulses that contain only a few cycles of the radiation field is briefly discussed.
\end{abstract}
\maketitle

\section{Introduction}
    The discovery of the quantum Hall effect is considered as a milestone in condensed matter physics that undeniably linked the topological structure of electronic wave functions to the macroscopic properties of a system\citep{Klitzing1980,Laughlin1981,Halperin1982a,Thouless1982}, ultimately leading to the description of a novel class of quantum states: Topological Chern or quantum Hall insulators (TIs or QHIs).~\citep{Hasan2010,Kane2011,Ando2013,Bernevig2013,Shen2013} The ground state of these non-interacting fermionic systems is well characterized by highly non-local order parameters, the Chern numbers associated to each Bloch band.  The celebrated bulk-boundary correspondence principle states that if the sum of these integer numbers up to the Fermi level is non-zero, gapless chiral states will be present at the edge of the system~\citep{Hasan2010,Bernevig2013}. The existence of these conducting boundary excitations in bulk-insulating materials leads to a manifold of quantum Hall signals, such as the quantum anomalous Hall effect~\citep{Haldane1988,Nagaosa2010} which is present even in the absence of external magnetic fields, a hallmark of non-trivial topology. 

Over the last decade, great advances in the field of anomalous Hall signals have been fuelled with the idea of engineering topological band structures by driving otherwise conventional materials with an external time-periodic potential~\citep{Oka2009,Oka2010,Kitagawa2010,Lindner2011,Zhou2011,Kitagawa2011,Perez-Piskunow2014,Usaj2014a}. The proposal of the so called Floquet Topological Insulators (FTI) opened the road for an external control of the properties of matter with the potentiality of optically turning on and off energy gaps~\citep{Calvo2011} containing chiral edge states in ultra-short time scales. Evidence of photon-dressed Floquet band structure has been revealed in time-resolved pump and probe spectroscopic measurements~\citep{Wang2013a}, but transport experiments with a Hall setup in these unique phases of matter are still yet to come. While the possibility to control topological transitions with light looks appealing, some of the concepts that are well established in unperturbed systems cannot be generalized in a simple way to this out of equilibrium phases. In fact, numerical approaches have shown that the zero magnetic field Hall conductance of stationary irradiated FTIs is not quantized~\citep{FoaTorres2014} nor related to the Chern number of the entire Floquet band~\citep{Perez-Piskunow2015}. 
\newline

Even more, the broader problem of dynamically reaching a topological regime when ramping an initially trivial Hamiltonian through a topological quantum phase transition and determinig what are the natural observables to look for is still a subject of ongoing discussion~\citep{DAlessio2015,Budich2016,Schler2017}. The time averaged Hall conductance following a quantum quench between two inequivalent topological phases was analyzed in several theoretical works~\citep{Dehghani2015,Dehghani2015b,Wang2016,Caio2016,Wilson2016,Schmitt2017}, unveiling that the final response is not necessarily quantized. A direct time-resolved evaluation of the bulk Hall current expectation value was also reported~\citep{Hu2016}, manifesting a non-trivial signal that builds up in time when making a controlled parameter ramp into a Chern insulator final Hamiltonian. Generally speaking, the well established bulk-edge correspondence in equilibrium is not guaranteed when dynamically preparing the topological phase. Discontinuities in bulk observables are due to the opening and closing of gaps in the instantaneous energy spectrum as the topological regime is reached, which unavoidably leads to a non-adiabatical population of the target states. 

In this work, we revisit some of these points by considering the development of a Hall response in isolated systems under coherent dynamics throughout a pulsed perturbation, motivated both by the theoretical understanding of the basic mechanisms that generate out of equilibrium quantum Hall signals and by current time-resolved experiments that are able to perform measurements during time periods shorter than characteristic relaxation timescales\citep{Higuchi2017}. On the other hand, mostly time averages that disregard dynamical features were reported in the literature. We address a simple toy model that switches a valley Hall signal in a honeycomb lattice and analyze its dynamics for different ramping protocols and as a function of doping. Furthermore, we consider the effect of irradiating the sample with a circularly-polarized infrared pump pulse that breaks time reversal symmetry and is expected to ramp the system into a Floquet topological phase. It is a relevant task to identify if experimentally accessible pump pulses with only a few femtosecond width and moderate frequencies are able to reveal some aspects of Floquet theory, even though the system is no longer in a stationary irradiated regime. It is also of interest to analyze the post-pulse response, in particular in the case of ultra-short pulses containing only a few cycles of the electromagnetic field. In the following sections we study and present results that will help clarifying all these and related points.
\section{The Model}
We start with a Hamiltonian describing the electronic structure of graphene and related 2D materials. 
\begin{eqnarray}
\nonumber
\mathcal{H}&=&-\sum_{\bm{k},s}\gamma\left[\phi_{\bm{k}}a^{\dagger}_{\bm{k}s}b^{}_{\bm{k}s}+\phi^{*}_{\bm{k}}b^{\dagger}_{\bm{k}s}a^{}_{\bm{k}s}\right]\\
&&+\sum_{\bm{k},s}\Delta \left[a^{\dagger}_{\bm{k}s} a^{}_{\bm{k}s} -b^{\dagger}_{\bm{k}s}b^{}_{\bm{k}s}\right]\,,
\label{eq1}
\end{eqnarray}
where $a^{}_{\bm{k}s}$ and $b^{}_{\bm{k}s}$ destroy an
electron with wavector $\bm{k}$ and spin $s$ in sublattices $A$ and $B$ of the honeycomb lattice, respectively. The matrix
element $\gamma$ corresponds to the nearest neighbor hopping and
\begin{equation}
\phi_{\bm{k}}=e^{iak_{y}}\left[1+2e^{-i\frac{3a}{2}k_{y}}\cos\left(\frac{a\sqrt{3}}{2}k_{x}\right)\right]\,,
\end{equation}
with $a$ the distance between neighboring sites. In the expression above, $\Delta$ is a mass like term that gaps the spectrum introducing a staggered on-site sublattice potential. This term breaks inversion symmetry and is then absent in graphene. In silicene or germanene, however, it can be induced by an electric field while in other 2D transition metal compounds it occurs naturally. From here on we drop the spin index keeping in mind that all states are double degenerate.

We consider a time-dependent perturbation that may break time-reversal (TR) symmetry but preserves translational symmetry. As a consequence the electron crystal momentum is conserved and the time dependent Hamiltonian has the form ${\cal{H}}(t)= \sum_{\bm{k}}{\cal{H}}_{\bm{k}}(t)$. The time dependence of ${\cal{H}}_{\bm{k}}(t)$ may be due to a time variation of the Hamiltonian parameters or to the action of a uniform circularly polarized electromagnetic field.

In this work we present analytical and numerical results using different techniques.  Analytical results are obtained using quasi-adiabatic or sudden approximations, Floquet theory for the case of time-periodic perturbations~\cite{Shirley1965,Sambe1973,Grifoni1998a,Kohler2005} and the two-time formalism for perturbations with two characteristic time scales.~\cite{Peskin1993}  The numerical results are obtained taking into account the full time evolution operator
\begin{equation}
U(t,t') = \prod_{\bm{k}} \mathcal{T} e^{-\frac{i}{\hbar}\int_{t^{\prime}} ^{t}\mathcal{H}_{\bm{k}}(t'')dt''} \,,
\end{equation}
where $\mathcal{T}$ is the time-ordering operator.
\section{The Hall Conductivity}
The calculation of the Hall conductivity under the effect of a time-dependent perturbation requires the use of out of equilibrium techniques. To describe this procedure we use linear response theory for the case of a time dependent Hamiltonian
\begin{equation}
\mathcal{H}_{\mathcal{V}}(t)=\mathcal{H}(t)+\mathcal{V}(t)\,,
\end{equation}
where $\mathcal{H}(t)$ is the Hamiltonian of the system including the time dependent perturbation and $\mathcal{V}(t)$ describes the action of the small bias. A generalized interaction representation for the wavefunctions $\ket{\psi_{I}(t)}$ and operators ${\cal{O}}_{I}(t)$ is defined as 
$\ket{\psi_{I}(t)} =U(-\infty,t)\ket{\psi_{S}(t)}=U(-\infty,t)U_{\mathcal{V}}(t,-\infty)\ket{\psi_{S}(-\infty)}$
and ${\cal{O}}_{I}(t)=U(-\infty,t){\cal{O}}_{S}(t)U(t,-\infty)$ where the subindex $S$ on the right hand side of these equations stands for Schr\"odinger representation and $U(t, t^{\prime})$ and $U_{\mathcal{V}}(t, t^{\prime})$ are the time evolution operators for the Hamiltonians $\mathcal{H}(t)$ and $\mathcal{H}_{\mathcal{V}}(t)$, respectively. Expanding $U_{\mathcal{V}}(t, t^{\prime})$ to first order in the small perturbation $\mathcal{V}(t)$ and assuming that at time $t=-\infty$ the system is in thermal equilibrium we obtain for our (non-interacting) system
\begin{eqnarray}
\langle\mathcal{O}_{S}(t)\rangle &=& \sum_{\alpha} f(\varepsilon_{\alpha})\Big(\!\bra{\psi_{\alpha}}\mathcal{O}_{I}(t)  \ket{\psi_{\alpha}}\\
\nonumber
\!&\!-\!&\!\frac{i}{\hbar}\!\int_{-\infty}^{\infty}\!dt^{\prime}\Theta(t-t')\bra{\psi_{\alpha}}[\mathcal{O}_{I}(t),\mathcal{V}_{I}(t^{\prime})]\ket{\psi_{\alpha}}\!\Big)
\end{eqnarray}
where $\ket{\psi_{\alpha}}$ and $\varepsilon_{\alpha}$ are the one-particle eigenfunctions and eigenvalues of $\mathcal{H}(t=-\infty)$,  $f(x)$ is the Fermi function and $[ \cdot\, , \cdot]$ indicates the commutator.

The Hall conductivity  is obtained from this expression for the case where $\mathcal{V}_{I}(t^{\prime})$ describes the effect of a bias field $E_0$ along, say,  the $y$-axis and ${\cal{O}}_{I}(t)$ is the current operator along the $x$-axis. The electric field in this case is described by a spatially homogeneous time-dependent vector potential  ${\bm{A}}_{b}(t)=-c E_0\frac{1}{\eta}\ln(1+e^{\eta t})\hat{\bm{y}}=-cE_0{\cal{W}}(t)\hat{\bm{y}}$ with $\eta>0$. Expanding the Hamiltonian to first order in $E_0$ we get 
\begin{equation}
{\cal{H}}_{\mathcal{V}}(t)={\cal{H}}(t)-\frac{1}{c} {\bm{A}}_{b}(t)\cdot\bm{j}\,,
\end{equation}
with $\bm{j}=-\frac{e}{\hbar} \sum_{\bm{k}}\nabla_{\bm k}{\cal{H}}(t)$. In terms of these quantities, the time dependent Hall conductivity for $t>>1/\eta$ is then given by
\begin{widetext}
\begin{eqnarray}
\nonumber
\sigma_{xy}(t)&=& -\frac{i}{\hbar}\sum_{\bm{k},\alpha} f(\varepsilon_{\bm{k}\alpha})
  \int_{-\infty}^{\infty}\Theta(t-t')\bra{\psi_{\bm{k}\alpha}}[j_{xI}(t),j_{yI}(t^{\prime})]  \ket{\psi_{\bm{k}\alpha}} {\cal{W}}(t^{\prime})dt^{\prime}\\
&=& -\frac{i}{\hbar}\sum_{\bm{k},\alpha} f(\varepsilon_{\bm{k}\alpha})
  \int_{-\infty}^{\infty}\Theta(t-t')\bra{\psi_{\bm{k}\alpha}}[U(-\infty,t)j_{x}U(t,-\infty),U(-\infty,t^{\prime})j_{y}U(t^{\prime},-\infty)]  \ket{\psi_{\bm{k}\alpha}} {\cal{W}}(t^{\prime})dt^{\prime} \,.
\label{eq3}
\end{eqnarray}
\end{widetext}
Here $\ket{\psi_{\bm{k}\alpha}}$ corresponds to an eigenstate of the system in equilibrium $(t=-\infty)$ with energy $\varepsilon_{\bm{k}\alpha}$ ($\alpha$ is the band index).
It is important to emphasize that in the case of perturbations of finite duration, such as pulses (see below), a  finite value for $\sigma_{xy}(t)$ after the perturbation should be understood as signaling the presence of a remanent Hall current.
\section{The Hall response of simple cases}
The above formulation of the Hall conductivity allows to calculate the Hall response for different models and conditions. In what follows we present numerical results and analytical approximations to interpret the behaviour of simple cases. For the subsequent analysis, it is useful to intoduce the quantum geometric tensor (also known as the Fubini-Study metric tensor of complex projective spaces~\citep{Study1905}), defined as
\begin{equation}
\mathcal{Q}_{\mu\nu}^{\alpha} := \langle D_{k_{\mu}}\psi_{\bm{k}\alpha}|D_{k_{\nu}}\psi_{\bm{k}\alpha}\rangle = \mathcal{G}_{\mu\nu}^{\alpha} + i\frac{\mathcal{F}_{\mu\nu}^{\alpha}}{2},
\label{FS}
\end{equation}
where $D_{k_{\mu}} = \partial_{k_{\mu}} - i\mathcal{A}_{\mu}$ is the covariant derivative and $\mathcal{A}_{\mu} = -i\langle\psi_{\bm{k}\alpha}|\partial_{k_{\mu}}\psi_{\bm{k}\alpha}\rangle$ the Berry connection. The imaginary part of $\mathcal{Q}_{\mu\nu}^{\alpha}$ is proportional to the widely known Berry curvature
\begin{equation}
\mathcal{F}^{\alpha}_{\mu\nu}(\bm{k}):= -i\bigg[\Bigg \langle \frac{\partial \psi_{\bm{k}\alpha}}{\partial k_{\mu}}\bigg|\frac{\partial \psi_{\bm{k}\alpha}}{\partial k_{\nu}}\Bigg \rangle-\Bigg \langle \frac{\partial \psi_{\bm{k}\alpha}}{\partial k_{\nu}}\bigg|\frac{\partial \psi_{\bm{k}\alpha}}{\partial k_{\mu}}\Bigg \rangle\bigg]\,,
\end{equation}
and the real part describes the metric that measures the distance between two nearby Bloch states \,\,\,\,\,\,\,\,\,\,\,\,\,\,\,\,\,\,\,\,\,\,\,\,\,\,\,\,\,\,\,\,\,\,\,\,\,\,\,\,\,\,\,\,\,\,\,\,\,\,\,\,\,\,\,$ds^2 = 1 - |\langle \psi_{\bm{k}\alpha}|\psi_{\bm{k}+d\bm{k},\alpha}\rangle|^2 = \sum_{\mu,\nu}\mathcal{G}^{\alpha}_{\mu \nu}dk_{\mu}dk_{\nu}$, with 
\begin{equation}
\begin{gathered}
\mathcal{G}_{\mu\nu}^{\alpha}(\bm{k}) := \frac{1}{2}\bigg[\Bigg \langle \frac{\partial \psi_{\bm{k}\alpha}}{\partial k_{\mu}}\bigg|\frac{\partial \psi_{\bm{k}\alpha}}{\partial k_{\nu}}\Bigg \rangle + \Bigg \langle \frac{\partial \psi_{\bm{k}\alpha}}{\partial k_{\nu}}\bigg|\frac{\partial \psi_{\bm{k}\alpha}}{\partial k_{\mu}}\Bigg \rangle\,\,\,\,\,\,\,\,\,\,\,\,\,\,\,\\
\!-\Bigg\langle\!\frac{\partial \psi_{\bm{k}\alpha}}{\partial k_{\mu}}\bigg|\psi_{\bm{k}\alpha}\!\Bigg \rangle\Bigg \langle\!\psi_{\bm{k}\alpha}\bigg|\frac{\partial \psi_{\bm{k}\alpha}}{\partial k_{\nu}}\!\Bigg \rangle\!-\!\Bigg \langle\! \frac{\partial \psi_{\bm{k}\alpha}}{\partial k_{\nu}}\bigg|\psi_{\bm{k}\alpha}\!\Bigg \rangle\!\Bigg \langle\psi_{\bm{k}\alpha}\bigg|\frac{\partial \psi_{\bm{k}\alpha}}{\partial k_{\mu}}\!\Bigg \rangle\!\bigg].
\end{gathered}
\end{equation}
\subsection{The equilibrium response}
The well known Hall conductivity of a system in equilibrium, with the bias field being the only external perturbation, represents a paradigm of the bulk-boundary correspondence: the topology of the band structure wave functions determines the number of current-carrying edge states. In this case, the time propagators in Eq. (\ref{eq3}) are simply given by $U_{0}(t_1,t_2 )=\prod_{\bm{k}}e^{-\frac{i}{\hbar}{\cal{H}}_{\bm{k}} (t_1-t_2)}$ and, after some algebra, the final result for the Hall conductivity of a two-band model can be written in terms of the aforementioned gauge invariant tensor [see Eq. (\ref{FS})]
\bea
\notag
\sigma_{xy} &=& \frac{e^{2}}{\hbar}\sum_{\bm{k}\alpha} f(\varepsilon_{\bm{k},\alpha})\Big[\mathcal{F}^{\alpha}_{xy}(\bm{k}) + t\frac{4\varepsilon_{\bm{k}\alpha}}{\hbar}\mathcal{G}^{\alpha}_{xy}(\bm{k})\Big]\\
&=& \frac{e^{2}}{h}\frac{1}{2\pi}\sum_{\alpha}\int_{\mathrm{BZ}} f(\varepsilon_{\bm{k}\alpha})\mathcal{F}^{\alpha}_{xy}(\bm{k})d^2{\bm{k}}.
\label{eq9}
\eea
The second (dangerously divergent) term in the first equality of Eq. (\ref{eq9}) integrates to zero in equipotentials over the Brillouin zone (BZ).

In the case of the Hamiltonian defined by Eq. (\ref{eq1}), the total Hall conductivity can be expressed as the sum of two contributions coming from states with wavector ${\bm{k}}$ close to the Dirac points $\bm{K}$ and $\bm{K}'$ of the BZ. Close to these points the Hamiltonian can be approximated by  
\begin{equation}
{\cal{H}}_{k}^{\xi}=\hbar v_F \bm{\sigma}\cdot\left(\xi k_x, k_y ,\frac{\Delta}{\hbar v_F} \right)\,,
\label{Hmass} 
\end{equation}
where now $\bm{k}=(k_{x},k_{y})$ is the wavevector measured from the $\bm{K}$ ($\xi=+$) or $\bm{K}'$ ($\xi=-$) points of the BZ, $v_{F}$ denotes the Fermi velocity and $\bm{\sigma}=(\sigma_x,\sigma_y,\sigma_z)$ are the Pauli matrices describing the pseudo-spin degree of freedom. 
For this particular case, the Berry curvature and the real part of the $x\text{-}y$ component of the metric reduce to
\bea
\notag
\mathcal{F}^{\alpha}_{xy} (\bm{k},\xi) &=& \pm\frac{\xi (\hbar v_F)^2 \Delta}{2 [(\hbar v_{F} k)^{2}+\Delta^{2}]^{3/2}}\\
\mathcal{G}^{\alpha}_{xy} (\bm{k},\xi) &=& \pm\frac{\xi (\hbar v_F k)^2 \sin(2\theta_{\bm{k}})}{4 [(\hbar v_{F} k)^{2}+{ \Delta}^{2}]^{2}}
\eea
where the $\pm$ signs correspond the conduction ($\alpha=c$) and valence ($\alpha=v$) bands respectively, and we have introduced the angle $\theta_{\bm{k}} = \text{tan}^{-1}(k_y/k_x)$. The Berry curvature has a non-trivial value due to the breaking of inversion symmetry produced by the mass-like term $\Delta$ in the Dirac hamiltonian. At zero temperature and for the Fermi energy $\varepsilon_{F}=0$ the quantized Hall conductivity can be expressed as
\begin{equation}
\sigma_{xy}^{\xi}= \frac{e^{2}}{h}{\cal{C}_{\xi}}\,,
\end{equation}
where  ${\cal{C}_{\xi}}$ is the contribution to the Chern number from states of the $\xi$ valley or Dirac cone,
\begin{equation}
{\cal{C}_{\xi}}=\frac{1}{2\pi}\int\mathcal{F}^{v}_{xy} (\bm{k},\xi)d^2\bm{k}=-\xi\frac{\text{sgn}(\Delta)}{2}\,.
\label{eq13}
\end{equation}
The total Chern ${\cal{C}}={\cal{C}_{+}}+{\cal{C}_{-}}=0$, indicating a zero Hall response characteristic of a trivial topology. However, as each Dirac cone gives a nonzero $\sigma_{xy}^{\xi}$, the electric field induces a transverse valley current: the valley quantum Hall effect.~\cite{Rycerz2007,Xiao2007,Sui2015}

\subsection{A simple model with a time dependent valley Hall signal.} 
As a reference situation, and for the sake of comparison with the more interesting case of radiation to be discused in the next section, here we summarize the results of a model in which the mass term $\Delta$ is turned on as $\Delta(t) = \Delta_{\infty}/(1 + e^{-\beta(t-t_0)})$. After a transient time it reaches a final value $\Delta_{\infty} = 50\,\text{meV}$ with the parameter $\beta$ controlling the velocity of the ramp. The Hamiltonian of the system for wavevectors near the Dirac points is given by Eq. (\ref{Hmass}) where now the mass term $\Delta$ acquires some time dependence. The dynamical response of the system depends on the way the mass is turned on, evolving from a quasi-adiabatic behavior for very slow ramps to quenched behavior for fast switchings.
\begin{figure}[b]
\includegraphics[width=\columnwidth]{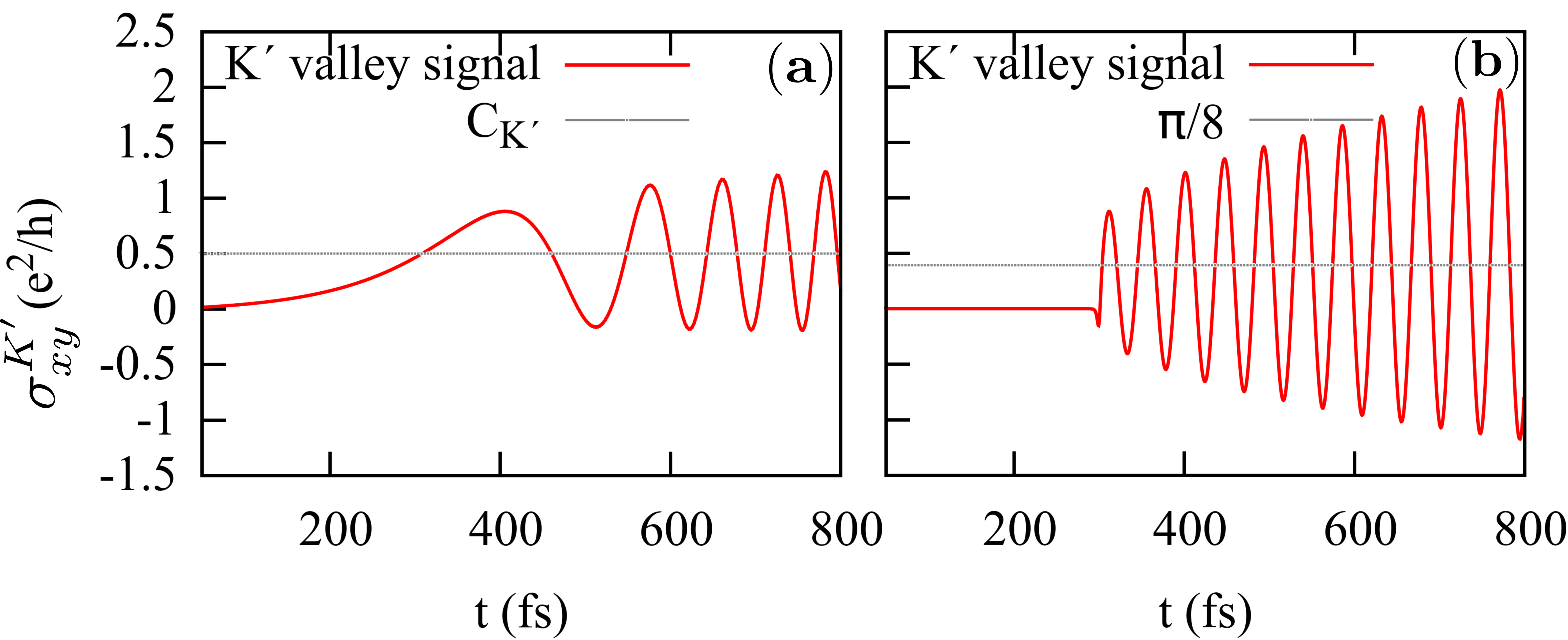}
\caption{Time resolved Hall conductivity at $\bm{K'}$ valley with a switch-on envelope $\Delta(t) = \Delta_{\infty}/(1 + e^{-\beta(t-t_0)})$. (\textbf{a}) $\beta = 0.01$ fs$^{-1}$ with the dashed horizontal line denoting the equilibrium expected value $C_{\bm{K'}}=0.5$;(\textbf{b}) $\beta = 1$ fs$^{-1}$ with the dashed horizontal line at the long-time limit for the sudden approximation $\overline{\sigma}_{xy}(\infty)=\pi/8$.}
\label{fig1}
\end{figure}

The numerical results for $\sigma_{xy}^{\bm{K'}}(t)$ are shown in Fig. \ref{fig1} for different parameters. For slow (quasi-adiabatic) switching-on of the mass term [Fig.~\ref{fig1}$\mathbf{(a)}$], the Hall conductivity increases while oscillating in time and the absolute value of its asymptotic average stays close to $\frac{e^{2}}{2h}$, the quantized expected response [see Eq.~(\ref{eq13})], shown with the dashed horizontal line. 

For a fast switching-on of the mass [Fig.~\ref{fig1}$\mathbf{(b)}$], $\sigma_{xy}^{\bm{K}'}(t)$ also shows oscillations with increasing amplitude and manifests an absolute value of the asymptotic time average smaller than $\frac{e^{2}}{2h}$. As shown in Appendix $\mathbf{A}$, the long-time limit of the mean Hall conductivity after a sudden parameter change is analytically found to be an integral of the Berry curvature of the target Hamiltonian $\mathcal{F}_{xy}^f(\bm{k})$ weighted by the occupations of the after-quench states $\ket{{\phi_{\bm{k}f}}}$, a result already obtained by Ref.~[\onlinecite{Wang2016}],
\begin{equation}
\overline{\sigma}_{xy}(\infty) = \frac{e^2}{\hbar}\!\sum_{\bm{k}\alpha}\sum_f f(\varepsilon_{\bm{k}\alpha})|\Lambda^{\bm{k}}_{f\alpha}|^2\mathcal{F}_{xy}^f(\bm{k}),
\end{equation}
with $\Lambda^{\bm{k}}_{f\alpha} = \langle{\phi_{\bm{k}f}}\ket{\psi_{\bm{k}\alpha}}$. In the particular case of a sudden turning-on of the mass at zero temperature, this asymptotic value is $\pi /8$, independent of the value of $\Delta_{\infty}$ (see Eq. \ref{A7}). 

After the transient time, the characteristic frequency of the valley Hall signal is given by the mass gap $2\Delta_{\infty}$ of the final Hamiltonian. The increasing amplitude of the oscillating response is due to the coherence terms in the wavefunctions, which have been dismissed in the diagonal ensemble used to calculate the mean value of the Hall conductance $\overline{\sigma}_{xy}(\infty)$. A linear increase in time of the amplitude is derived in Appendix $\bm{A}$ (see Eq.~\ref{eqA8}). A qualitative similar dependence can be seen to be present in the numerical results obtained in Ref.~[\onlinecite{Hu2016}] with a parameter ramp of the BHZ Hamiltonian~\citep{Qi2008}, indicating that this effect is more general than the particular model addressed in our work.

\begin{figure}[t]
\includegraphics[width=\columnwidth]{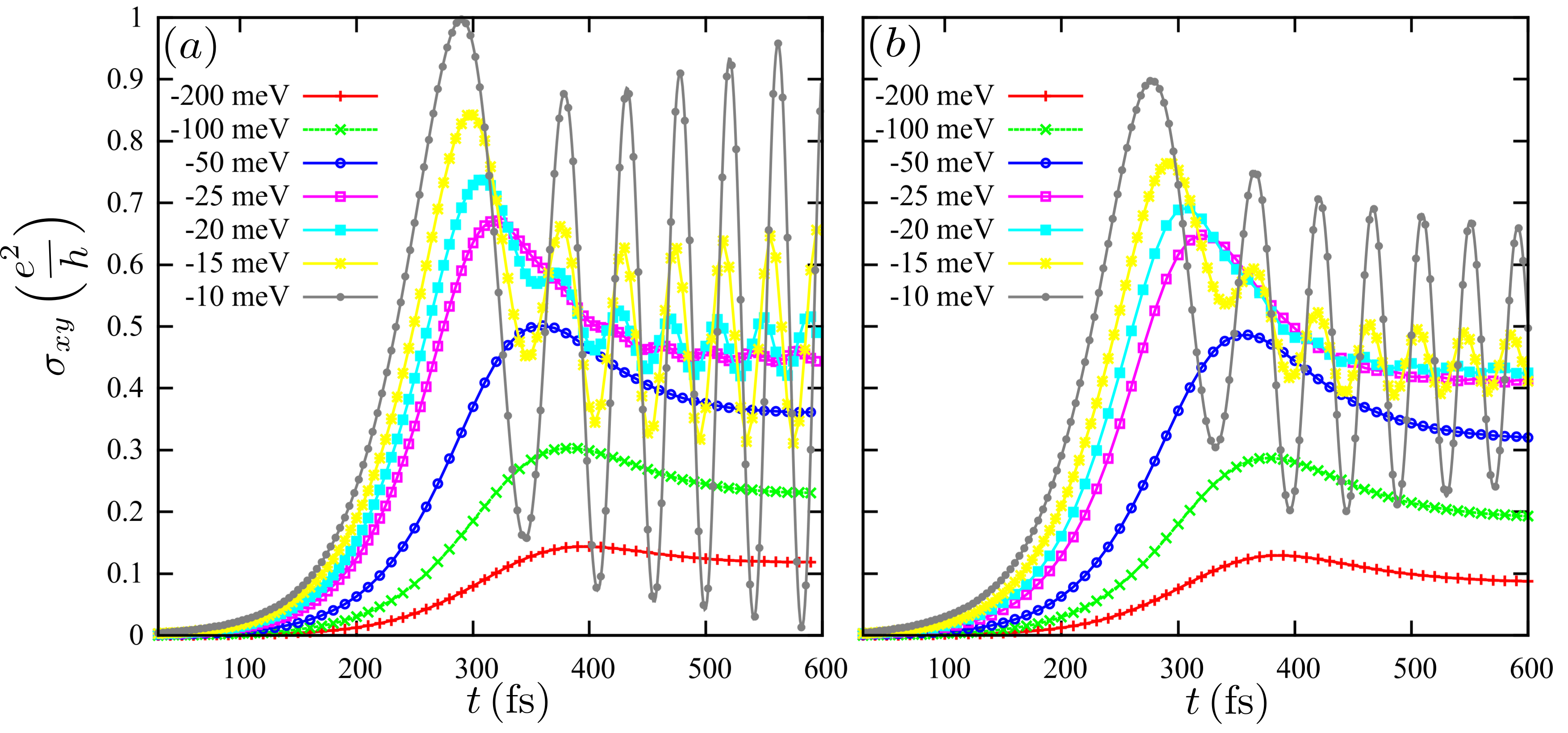}
\caption{Time resolved Hall conductivity at $\bm{K'}$ valley with a switch-on envelope of the mass term in Eq.~(\ref{eq3}) as $\Delta(t) = \Delta_{\infty}/(1+e^{-\beta(t-t_0)})$ with $\beta = 0.01\,\text{1/fs}$. Different curves are obtained by changing the Fermi level. (a) Obtained with the exact time evolution of the Bloch states and (b) with the time evolution dictated by the total adiabatic hypothesis.}
\label{fig_flevels}
\end{figure}

In Fig.~\ref{fig_flevels} we show the time resolved Hall conductivity calculated as a function of doping. Fig.~\ref{fig_flevels}(a) was obtained with the full time evolution of the Bloch states and (b) with a total adiabatic evolution. For $\varepsilon_F \lesssim-\Delta_\infty/2$ both are comparable. The long-time mean value is diminished respect to the quantized value $e^2/2h$ on account of a reduced contribution of the total Berry curvature to the final response when considering the occupied states. When the Fermi level gets closer to the Dirac point, oscillations with the effective gap $2\sqrt{\Delta_{\infty}^2 + \varepsilon_F^2}$ can be appreciated. In the case of the total adiabatic approximation [Fig.~\ref{fig_flevels}(b)], these oscillations damp out for sufficiently long times (see  Eq.~(\ref{A3}) in Appendix $\mathbf{A}$), since the wavefunction remains a pure state in the lower instantaneous band and no coherent terms are allowed during the evolution.
\section{Radiation Driven System: Floquet picture and two-time dynamics.}
This section contains the central results of our work. Here we consider a uniform circularly polarized electromagnetic field described by the vector potential $\bm{A}(t)=\Re [A_{0}(t) e^{-\mathrm{i}\Omega t}(\hat{\bm{x}}+\ci\,\hat{\bm{y}})]$. The electrons-radiation coupling is described through the Peierls substitution that in our case reduces to replace $\phi(\bm{k})$ by $\phi(\bm{k}+ \frac{e}{\hbar c}\bm{A}(t))$ with $e$ the absolute value of the electron charge and $c$ the speed of light.

The uniform electromagnetic field, describing a plane wave with its wavector perpendicular to the plane of the sample, breaks time-reversal (TR) symmetry and preserves translational symmetry. The TR symmetry breaking by the circularly polarized field is important to generate non-trivial topological properties~\cite{Oka2009,Rudner2013} and Floquet chiral edge states~\cite{Perez-Piskunow2014}.
For frequencies $\Omega$ on the infrared side of the spectrum, all the physics takes place at low energies, i.e. close to the Dirac points.  The time dependent Hamiltonian for each wavevector $\bm{k}$ around the Dirac points of the BZ is now given by
\begin{equation}
{\cal{H}}_{\bm{k}}^{\xi}(t)=\hbar v_F \bm{\sigma}\cdot\left[\xi\left( k_x+\frac{e}{\hbar c}A_{x}(t)\right), k_y+\frac{e}{\hbar c}A_{y}(t) ,\frac{\Delta}{\hbar v_F} \right]\,.
\label{eq14}
\end{equation}
For a constant amplitude of the radiation field $A_{0}(t)=A_{0}$ the Hamiltonian is time-periodic with period $T=2\pi/\Omega$. In this case, the set of solutions of the Schr\"odinger equation can be expressed within the Floquet formalism, which states that the wavefunctions have the form $\ket{\psi_{\alpha}(t)}=\exp(-\ci\varepsilon_{\alpha}t/\hbar)\ket{\phi_{\alpha}(t)}$ where $\ket{\phi_{\alpha}(t)}$ are known as the Floquet modes, with the same time-periodicity as the Hamiltonian $\ket{\phi_{\alpha}(t+T)}=\ket{\phi_\alpha(t)}$ and $\varepsilon_{\alpha}$ as the quasi-energies~\cite{Shirley1965,Grifoni1998a,Sambe1973}. The Floquet modes $\ket{\phi_{\alpha}}$ are eigenfunctions of the Floquet operator $\hat{\cal{H}}_F = {\cal{H}}-i\hbar \frac{\partial}{\partial t}$  with eigenvalues $\varepsilon_{\alpha}$
\begin{equation}
\hat{\cal{H}}_F\ket{\phi_{\alpha}(t)} =\varepsilon_{\alpha}\ket{\phi_{\alpha}(t)}\,.
\label{floquetEq}
\end{equation}
\begin{figure}[t]
\includegraphics[width=1.1\columnwidth]{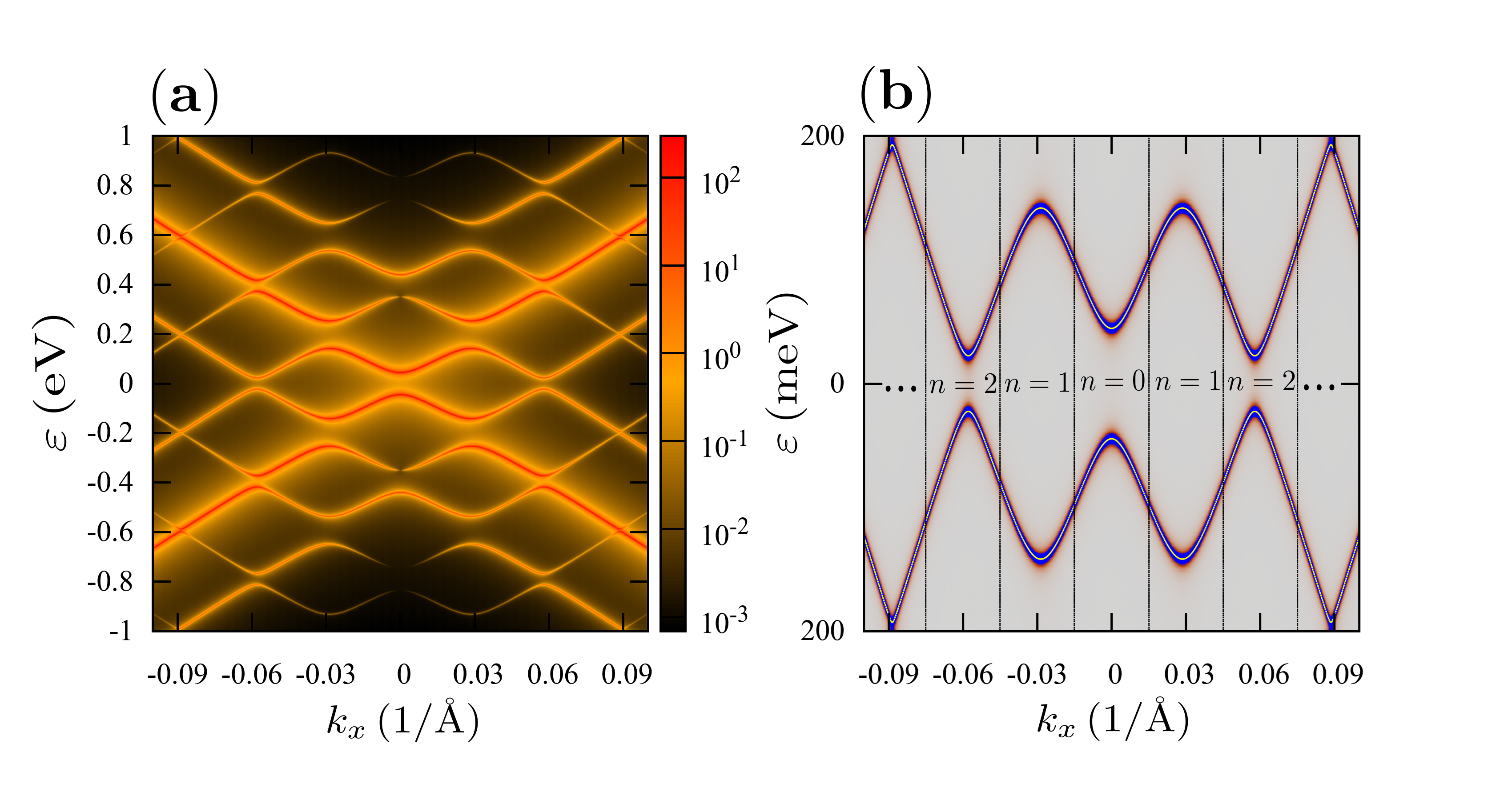}
\caption{$\mathbf{(a)}$ Floquet band structure of the Hamiltonian (\ref{eq14}) with circularly polarized laser in the extended Floquet scheme weighted by the density of states projected onto the replica $m=0$. $\mathbf{(b)}$ Floquet bands in the FFZ, the gaps are at the zone centre $\varepsilon = 0$ or boundary $\varepsilon = \hbar\Omega/2$. Vertical lines show the chosen regions to analyze the different contributions to the Hall response $\sigma_{xy}^{[n]}$ with $n=0,1,2,...$ [see Eq. (\ref{eq21})].}
\label{fig3}
\end{figure}

Decomposing the periodic modes in a Fourier basis, Eq. (\ref{floquetEq}) reduces to an eigenvalue problem in the composed Sambe space ${\cal{R}}\otimes{\mathcal{L}_2(0,T)}$, where $\cal{R}$ is the usual Hilbert space and $\mathcal{L}_2(0,T)$ is the space of periodic square integrable functions with period $T$. With the periodic functions spanned in a set of orthonormal functions $e^{i m\Omega t}$, the Floquet operator is now a time-independent infinite hamiltonian ${\cal{H}}_ F^{\infty}$. If the eigenvectors of ${\cal{H}}_ F^{\infty}$ are $(\cdots, u_{m,\alpha}, \cdots,u_{1,\alpha } , u_{0,\alpha } , u_{-1,\alpha } ,\cdots)^ {T}$, the time dependent wavefunctions have the form ${\ket{\Psi_{\alpha}(t)}}=e^{-i\varepsilon_{\alpha} t} \sum_{m} e^{im\Omega t} u_{m,\alpha}$ where the quasi-energy can be taken in the first Floquet zone (FFZ), that is $-\frac{\hbar \Omega}{2}<\varepsilon_{\alpha}\leq \frac{\hbar \Omega}{2}$. These wavefunctions describe coherent superposition of electronic and photonic states.

The time averaged band structure  is shown in Fig. \ref{fig3}$\mathbf{(a)}$ in the extended FZ scheme ($\varepsilon+m \hbar \Omega$) where the gaps at energies $ m \hbar \Omega /2 $ are apparent. In the FFZ [Fig. \ref{fig3}$\mathbf{(b)}$] the Floquet bands show the gaps at the zone centre $\varepsilon=0$ or at the zone boundary $ \varepsilon=\pm \hbar\Omega /2$. In a graphene strip, all these gaps are bridged by protected edge states~\citep{Perez-Piskunow2015}.The intensity in the color map of the figure indicates the contribution of each band to the time average of the density of states.     

If the amplitude of the radiation field $A_{0}(t)$ has some time dependence, breaking the time periodicity of the Hamiltonian, Floquet theory doesn't apply. In this case, it is useful to resort to a mathematical formulation of the evolution equation in an extended Hilbert space~\cite{Pfeifer1983,Breuer1989,Peskin1993,Drese1999}
\begin{equation}
[\mathcal{H}(\tau,t) - i\hbar\partial_t]\Ket{\psi(\tau,t)} = i\hbar\partial_{\tau}\Ket{\psi(\tau,t)},
\label{eq2t}
\end{equation}
where two time variables $t$ and $\tau$ are introduced. The former will be associated to the fast time-periodic evolution, while the later is intended to account for the slow variation of the pulse shape $A_0(\tau)$. The
two-time wavefunctions $\Ket{\psi(\tau,t)}$ belong to a generalized Hilbert space with an inner product defined as
\begin{equation}
\langle\langle \psi_{\alpha}(\tau,t)\Ket{\psi_{\beta}(\tau,t)} = \frac{1}{T}\int_{0}^{T}\langle\psi_{\alpha}(\tau,t)\ket{\psi_{\beta}(\tau,t)}dt.
\end{equation}
Restricting the solution to the contour $\tau = t$ it is possible to reobtain the physical wavefunction that satisfies the original Schr\"odinger equation~\cite{Pfeifer1983}
\begin{equation}
\Ket{\psi(\tau,t)}\Big\rvert_{\tau = t} = \ket{\psi(t)}.
\end{equation}
The advantage of this formalism manifests itself in the fact that the evolution equation distinguishes two different time scales, preserving the $2\pi/\Omega$ periodicity in the fast time coordinate and also taking into consideration the pulse modulation. In fact, Floquet wavefunctions are the adiabatic solutions of Eq. (\ref{eq2t}), since they obey an eigenvalue equation for the instantaneous Floquet operator 
\begin{equation}
\hat{\mathcal{H}}_F(\tau,t)\Ket{\phi_{\alpha}(\tau,t)} = \varepsilon_{\alpha}(\tau)\Ket{\phi_{\alpha}(\tau,t)},
\end{equation}
with $\varepsilon_{\alpha}(\tau)$ the instantaneous quasi-energies at time $\tau$.
A formal solution for the two-time wavevector could be achieved when it is decomposed in the Fourier basis $\Ket{\psi(\tau,t)} = \sum_m e^{im\Omega t}\ket{\chi_m(\tau)}$. Expanding Eq. (\ref{eq2t}) in it's normal modes, the state vector $\bar{\chi}(\tau) = (\cdots, \chi_{m}(\tau), \cdots,\chi_{1}(\tau) , \chi_{0}(\tau), \chi_{-1}(\tau),\cdots)^ {T}$ satisfies a slow time-dependent Floquet-Schr\"odinger equation
\begin{equation}
\mathcal{H}_F^{\infty}(\tau)\bar{\chi}(\tau)= i\hbar\partial_{\tau}\bar{\chi}(\tau),
\label{floquetscrho}
\end{equation}
where $\mathcal{H}_F^{\infty}(\tau)$ has block components defined by $H^{n,m}_{F}(\tau) = \frac{1}{T}\int_0^{T}dt\, e^{i(n-m)\Omega t} H(\tau,t) + n\hbar\Omega\delta_{n,m}$. If the switching of the electromagnetic field were to be considered completely adiabatic in the whole Brillouin zone, the Floquet basis would be accurate for the description of the evolved states. In fact, for photon energies $\hbar\Omega$ larger than the bandwidth, the FFZ is such that it contains the entire unperturbed band. Hence, for an undoped  system, all negative quasienergies are occupied and the positive ones are empty, making the Dirac points the only relevant gap for the adiabatic Floquet dynamics:  this is the so called \textit{non-resonant regime}.  On the contrary, when $\hbar\Omega$ is smaller than the bandwidth, intraband resonances between valence and conduction states occur, breaking the Floquet adiabatic picture. The Floquet `ground-state' does not coincide with the initial Slater determinant for any choice of the Floquet BZ~\citep{Privitera2016} and the lifting of degeneracies in the instantaneous Floquet spectrum at states $\bm{k}$ resonant with the radiation field generates tunneling between replicas resulting in an involved dynamical response.
\begin{figure}[t]
\includegraphics[width=0.9\columnwidth]{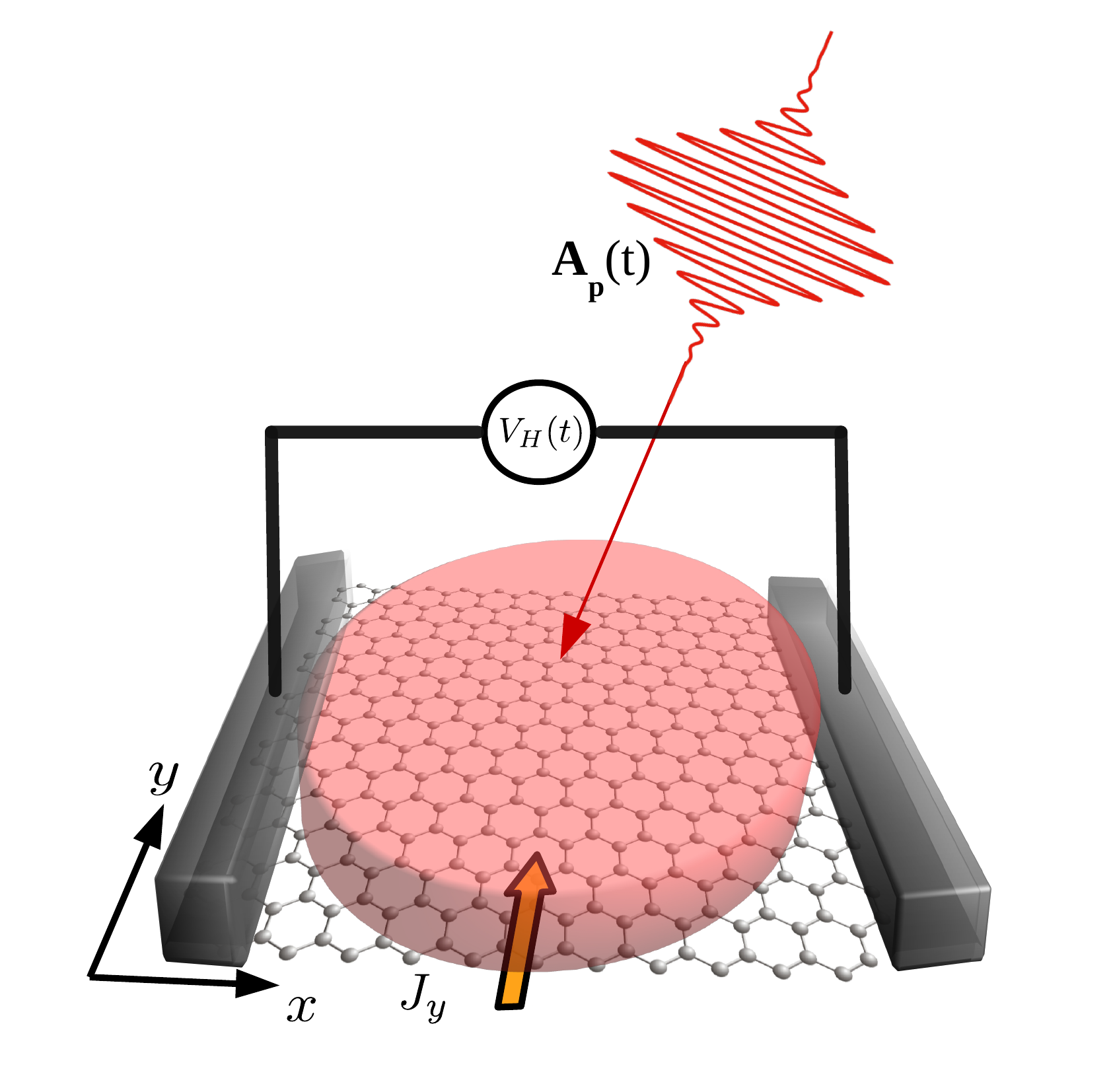}
\caption{Schematic representation of a transport measurement experiment of the Hall conductivity. Both a dc bias induced current $J_y$ in the $\hat{y}$ direction and a short pulse, characterized by a vector potential $A_p(t)$, are applied. We will assume the sample is irradiated with a spatially homogeneous electric field.}
\label{fig4}
\end{figure}
In what follows we formulate the problem of pulses of circularly polarized radiation with frequency $\Omega$ and a Gaussian envelope $A_0(t)$. A schematic representation of an experimental setup for measuring the Hall response of the bulk system is shown in Fig. \ref{fig4}.

\begin{figure}[t]
\includegraphics[width=\columnwidth]{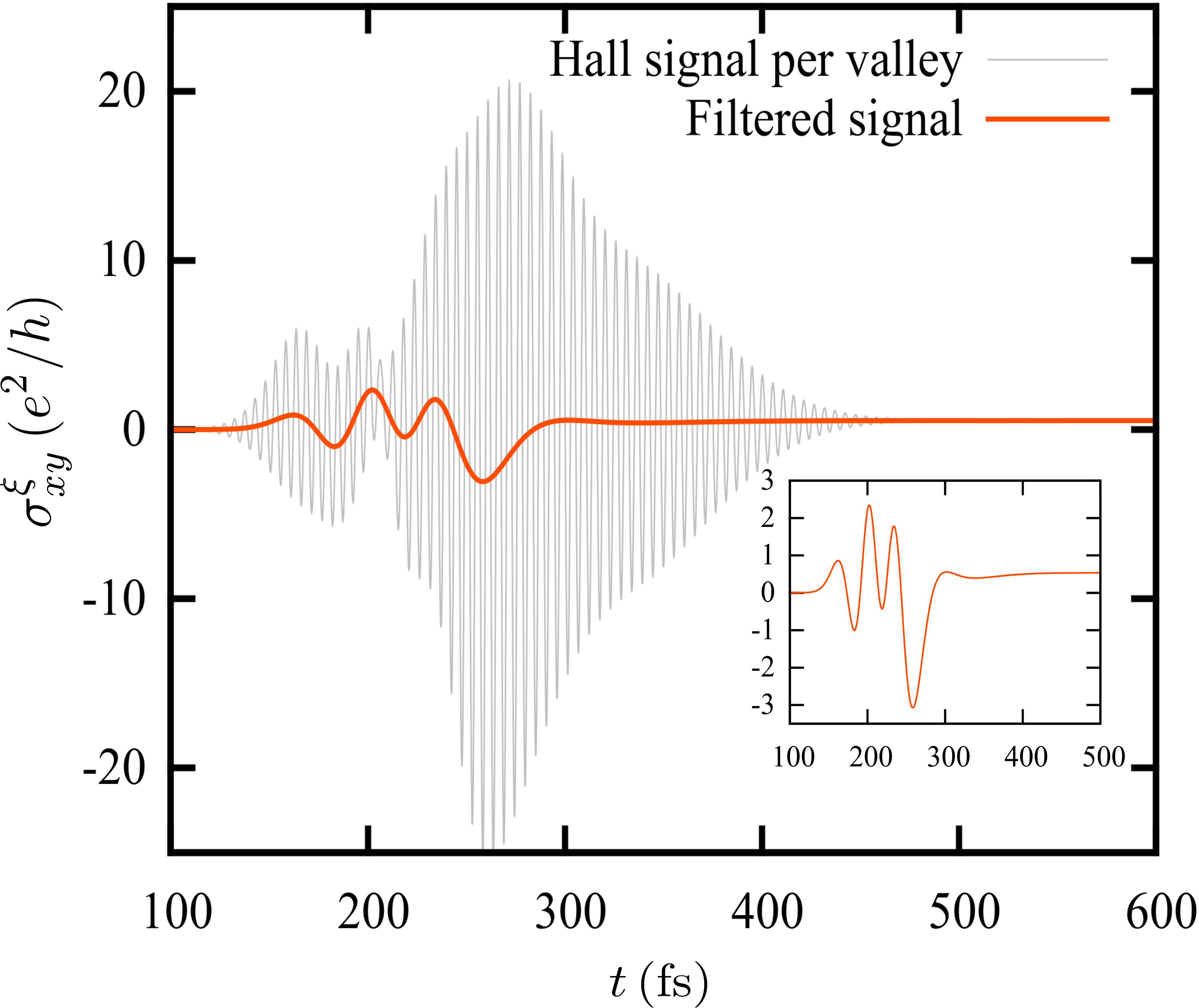}
\caption{Time resolved Hall conductivity per valley $\sigma_{xy}^{\xi}(t)$. In the inset we show the signal with the fast oscillations of frequency $2\Omega$ filtered.}
\label{fig5}
\end{figure}

 In Fig. \ref{fig5} we present  results for a pulse with photon energy of $\hbar\Omega = 400\,\text{meV}$, the gaussian width of the pulse is of $50\,\text{fs}$ and its maximum amplitude of $e v_f A_0 = 140\,\text{meV}$. The chemical potencial has been taken at the Dirac Point $\varepsilon_F=0$. The frequency of the electromagnetic field is taken to be smaller than the bandwidth of the unperturbed Bloch bands, since this resonant regime is more likely to be experimentally feasible. The result can be summarized as follows:
\begin{enumerate}[(i)]
\item Before the pulse the total conductivity ($\sum_{\xi}\sigma_{xy}^{\xi}$) is zero, after the pulse it converges to a finite value (representing a remanent Hall current)
\item During the pulse a high frequency oscillation of $2\Omega$ is observed.
\item Once the fast oscillations are filtered, the signal shows some small frequency oscillations during the pulse and remains constant after the pulse. In the figure, the filtered signal is obtained by Fourier transforming, eliminating the high frequency components and transforming back to time domain.
\end{enumerate}

The fast $2\Omega$ oscillations are a special case of the heterodyining effect, characteristic of periodically driven systems~\cite{Oka2016}. In our case, this is a consequence of a symmetry of the model. In fact, when considering the radiation amplitude constant within a period, the Hamiltonian [see Eq. (\ref{eq14})] is invariant under the operation $\mathbf{T}_{\Omega}(\theta_{\bm{k}})\mathbf{R}(\theta_{\bm{k}})$, where $\mathbf{R}(\theta_{\bm{k}})$  is a rotation in reciprocal space around the $z$-axis and $\mathbf{T}_{\Omega}(\theta_{\bm{k}})$ is a time translation that rotates the phase of the circularly polarized electric field by changing $t\rightarrow t + \theta_{\bm{k}}/\Omega$. By considering this symmetry operation in the Kubo formula, a simple angle integration shows that the only high-frequency mode in the response is the one with twice the driving frequency. This selection rule is demonstrated carefully in Appendix $\mathbf{B}$, explaining the supression of other multiples of the driving frequency.

\begin{figure}[t]
\includegraphics[width=\columnwidth]{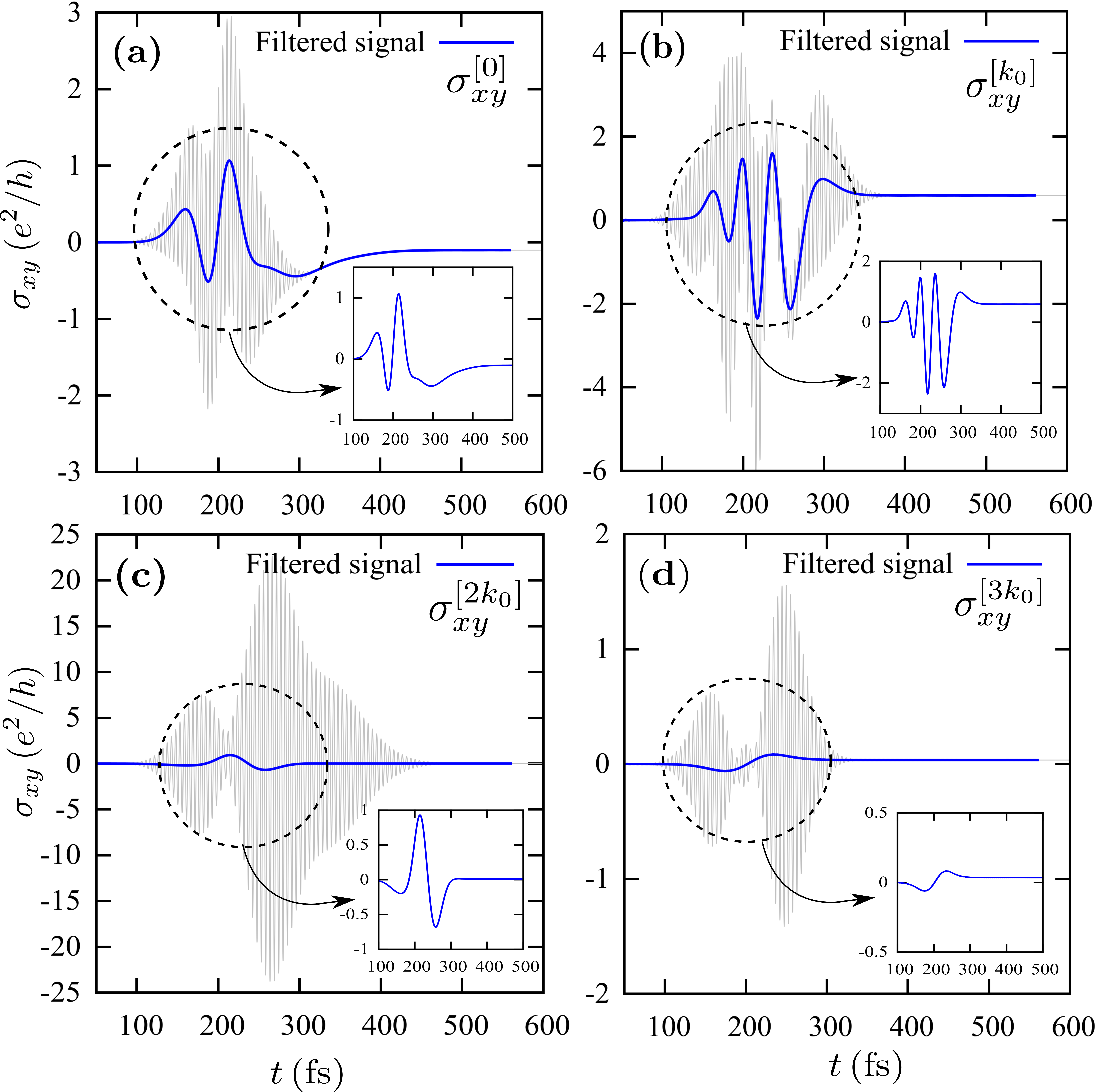}
\caption{Time resolved contributions to the Hall conductivity per valley $\sigma_{xy}^{[nk_0]}(t)$ coming from different Floquet gaps in the BZ, with its corresponding filtered signal. (a) $n=0$, (b) $n=1$, (c) $n=2$ and $n=3$.}
\label{fig6}
\end{figure}

In order to interpret the low frequency behaviour we analyze the contribution of states with different wavevectors $\bm{k}$. To this end we define
\begin{eqnarray}
\notag
\sigma_{xy}^{[0]}(t) &=& \int_{0}^{\frac{k_0}{2}}\int_{0}^{2\pi}\sigma_{xy}(\bm{k},t) k dk d\theta_{\bm{k}}\,,\\
\sigma_{xy}^{[nk_0]}(t) &=& \int_{\frac{(2n-1)k_0}{2}}^{\frac{(2n+1)k_0}{2}}\int_{0}^{2\pi}\sigma_{xy}(\bm{k},t) k dk d\theta_{\bm{k}},
\label{eq21}
\end{eqnarray}
where $k_0 = \frac{\Omega}{2 v_F}$ and $n$ a natural number. This partial integration is performed in order to analyze separatly the contribution of each of the resonances $n k_0$ where Floquet gaps occur. The result is show in Fig. (\ref{fig6}).
It's interesting to note that states near \textit{all} of the gaps generated by few photon processes manifest a non trivial response, not only those that cross the Fermi energy. With this choice of parameters the system is in a resonant regime, where a complex redistribution of electronic occupation of states takes place during the pulse. The small frequency oscillations are in each case in correspondance with the local instantaneous gap generated in the Floquet spectrum, which unmasks the fact that the wavefunction behaves as a coherent superposition of Floquet states. The mean and after-pulse value of the Hall response are highly dependent on the pump envelope: we are far from the limit of a clear quantized Hall regime, which could only be achieved if an ideal adiabatic population of Floquet modes takes place. Even if the hamiltonian after the pulse is in its topological trivial form, there's still a Hall current due to the proliferation of electron-hole pairs created throughout the excitation. 

The contribution coming from the Dirac points $\sigma_{xy}^{[0]}$ can be understood within the simple model exposed in the previous section: a mass-like switching term in the Dirac Hamiltonian. The two-time formalism provides an effective slow-time evolution for states near those gaps, as shown in Appendix $\mathbf{C}$. It can be shown that to second order in $\eta(t) = e v_f A(t)/\hbar \Omega$ and taking the limit $\dot{\eta}(t)\rightarrow 0$, the state vector can be approximated by

\begin{eqnarray}
|\psi_{\xi}(\tau,t)\rangle\rangle\Big\rvert_{\tau = t}\!&=&\!\Big[\mathcal{I}^{2\times 2}\Big(1\!-\!\frac{\eta^2(t)}{2}\Big)\\
\nonumber
\!&+&\!\eta(t)[\sigma_{+}(\xi)e^{-i\Omega t}\!-\!\sigma_{-}(\xi)e^{i\Omega t}]\Big]\widetilde{\chi}_0(\xi,t),
\label{aprox_wf}
\end{eqnarray}
where
\begin{equation}
\widetilde{\chi}_0(\xi,t) = U^{\xi}_{\text{eff}}(t,-\infty)\ket{\psi_{\bm{k}v\xi}}
\end{equation}
with the evolution operator corresponing to the Hamiltonian from Eq. (\ref{Hmass}) but with a valley-dependent mass term 
\begin{equation}
\Delta(\xi,\tau) = -\xi\frac{[e v_f A(\tau)]^2}{c^2\hbar\Omega}.
\label{eqvdpm}
\end{equation}
Using Eq. (\ref{aprox_wf}) to calculate the  transverse Hall response we find that the terms with the filtered fast oscillations come from the current-current correlation function $[\xi\sigma_x^{\text{eff}}(t),\sigma_y^{\text{eff}}(t')]$, where we define the operators in the effective interaction picture as $\sigma_{x,y}^{\text{eff}}(t) = U_{\text{eff}}(-\infty,t)\sigma_{x,y}U_{\text{eff}}(t,-\infty)$.
\begin{figure}[t]
\includegraphics[width=\columnwidth]{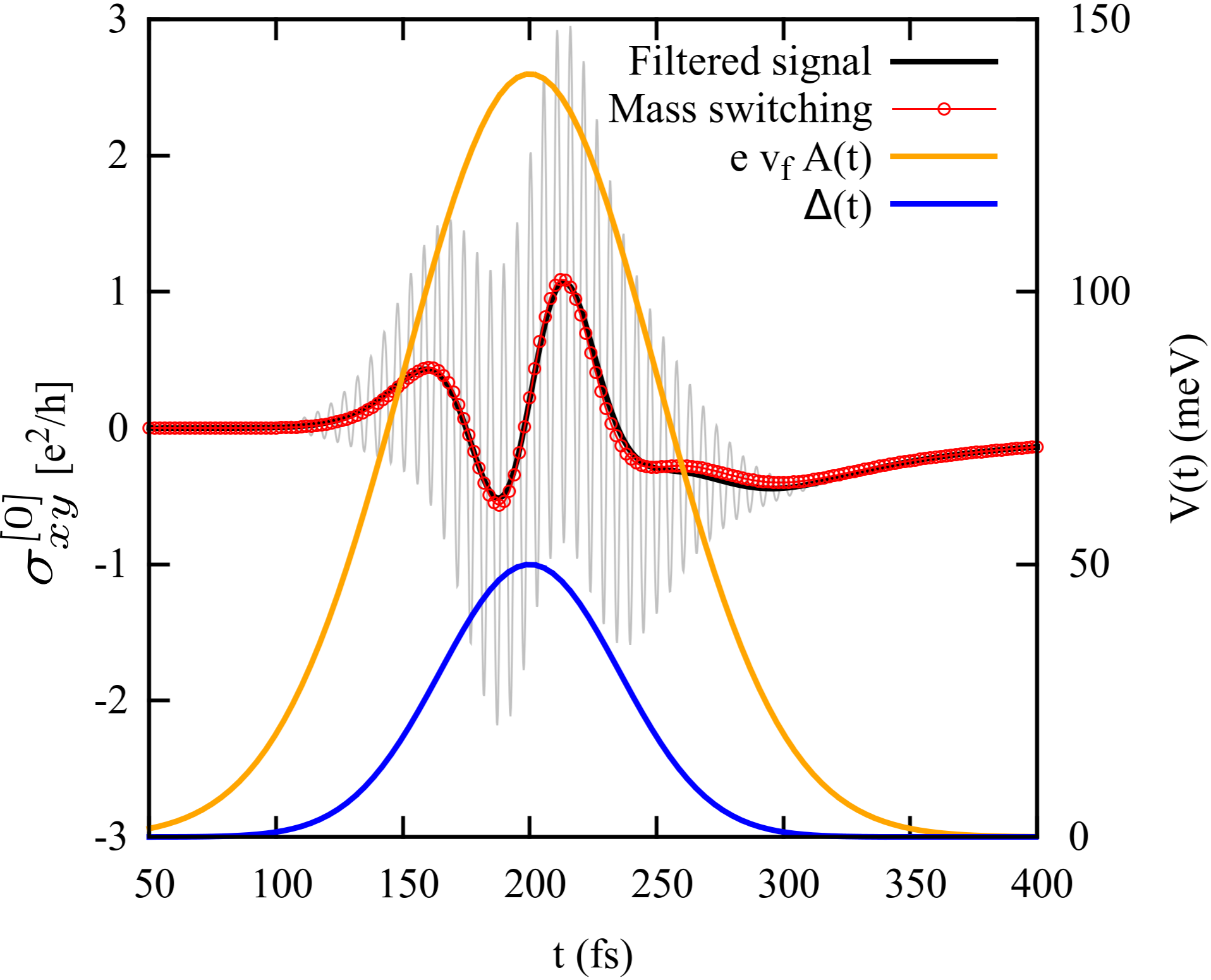}
\caption{Contribution to the Hall conductivity per valley coming from the Dirac cones, its corresponding filtered signal and the comparison with the mass-switching model. Both the pump envelope $e v_f A(\tau)$ and the pulsed mass perturbation $\Delta(\xi,\tau) = -\xi\frac{[e v_f A(\tau)]^2}{c^2\hbar\Omega}$ are shown.}
\label{fig7}
\end{figure} 
The comparison between the model and the numerical result is shown in Fig.~\ref{fig7}, finding a good agreement between both of them. The pump envelope and the effective mass perturbation are plotted, the two being related through Eq.~(\ref{eqvdpm}).

Interestingly, in the case of large frequencies, where the only contribution to the Hall conductance is expected to come from the Dirac points, this simple model explains analitically some numerical results already obtained in Ref.~[\onlinecite{Dehghani2015}]. In this work they found that after a sudden quench of the radiation field, the transverse conductivity converged to a finite value while increasing $\Omega$. Within our model, a simple calculation of the response [see Eq. (\ref{A7}) in Appendix $\mathbf{A}$] yields
\begin{equation}
\sum_{\xi}\bar{\sigma}^\xi_{xy}(\infty)=-\sum_{\xi}\xi\frac{e^2}{h}\frac{\pi}{8}\text{sgn}(\Delta(\xi)) = \frac{e^2}{h}\frac{\pi}{4},
\end{equation}
which seems to be in agreement with their work. We can also understand that ramps with lower velocities can in principle make this value approach to the expected topological quantized result, since a slower switch-on protocol [like the one in Fig. \ref{fig1}$\mathbf{(a)}$] would adiabatically populate the Floquet states near these valleys. Numerical results confirm this tendency.

No simple analytical approaches have been found to describe the low frequency signal coming from the rest of the dynamical gaps, since there is not an effective slow time evolution that can be disentangled from the high-frequency oscillations for such resonant cases~\citep{Novienko2017}.

\begin{figure}[t]
\includegraphics[width=\columnwidth]{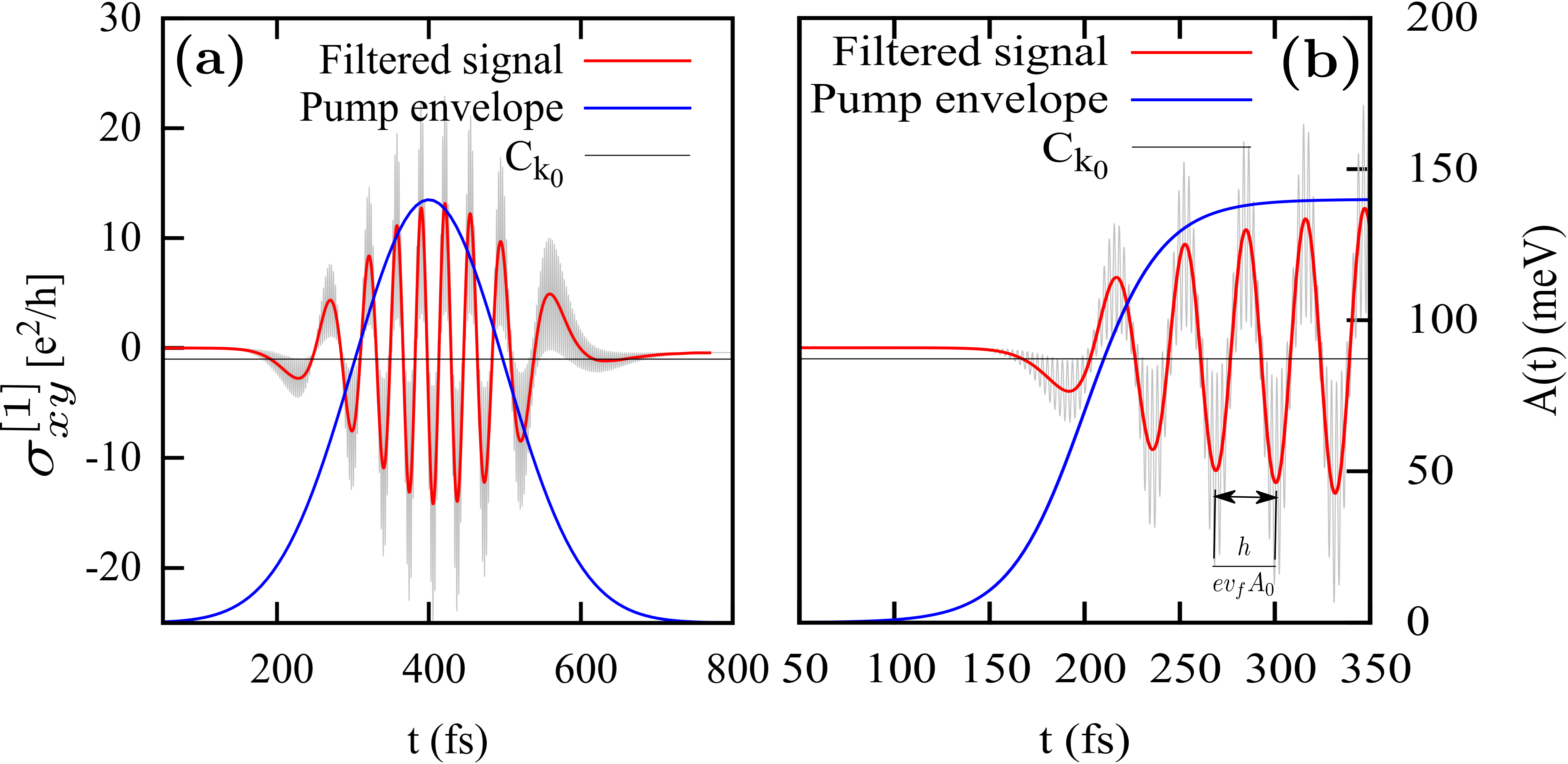}
\caption{Contribution to the Hall conductivity per valley $\sigma_{xy}^{[k_0]}(t)$ coming from states $\bm{k}$ near $k_0 = \frac{\Omega}{2 v_f}$, resonant with the photon energy $\hbar\Omega = 800\,$meV. We indicate its corresponding filtered signal and use for the sake of comparison two different pump switch-on protocols with \textbf{(a)} being Gaussian and \textbf{(b)} one that reaches a final value of $e v_f A_0 = 140\,$meV after a transient time. The horizontal line stands for the expected value in the case of adiabatic population of Floquet bands, the contribution~\cite{Perez-Piskunow2015} to the Chern number at each valley: $C_{k_0} = -1$.}
\label{fig8}
\end{figure}

Results obtained with the chemical potential at the first dynamical gap--the Floquet zone boundary--and a higher driving frequency, taken to be $\hbar\Omega = 800\,$meV, are shown in Fig. \ref{fig8}. In this case the main contribution to the Hall conductance comes from states $\bm{k}$ resonant with the photon energy. In Fig. \ref{fig8}\textbf{(a)} the pump envelope is chosen to be Gaussian while in Fig. \ref{fig8}\textbf{(b)} it reaches a final value after a transient time (the blue line shows its profile). In the latter case, the low frequency oscillations are well defined by the Floquet gap calculated at the lowest order, which is linear with the amplitude of the radiation field and independent of the frequency of the driving. As can be seen in the figure, the mean response at the center of the pulse and at large times for the switch-on case approaches the quantize value $C_{k_0} = -e^2/h$. This is consistent with the number of edge states bridging the gap of a finite samples with significat weight on the $m=0$ Floquet replica of the extended zone scheme.~\cite{Perez-Piskunow2015} Also, the sign difference between the mean response in the dynamical and non-resonant Dirac gap follows the fact that the chiral edge states have opposite velocities at each gap. These features are consistent with those observed in mumerical calculations of the Hall conductance in finite samples with non-irradiated leads.~\cite{FoaTorres2014} 

For the case of a sudden switch-on of the time dependent perturbation an approximate expression for the asymptotic long-time average Hall conductivity, like Eq.~(\ref{A4}) in Appendix $\mathbf{A}$, has also been obtained for the case of Floquet systems~\cite{Dehghani2015b}. These approximations give a compact and simple form for the long-time behavior of the Hall signal averaged over a period of the driving field in which the Berry curvature is weighted by the projection of the final state on the eigenstates of the Floquet Hamiltonian. It can be shown that the time average Hall conductivity given by such approximate expression, valid for the case of any saturating perturbation with a corresponding redefinition of the bands population, cannot exceed the absolute value of $\frac{1}{2}C_{k_0}e^2/h$ at each valley in the dynamical gap $k_0$. Consequently, this can at most describe the case of the undoped system and fails for the resonant case with $\varepsilon_F=-\hbar\Omega/2$. Note that such compact expressions are obtained by neglecting the off-diagonal terms describing inter-band transitions. If the Fermi energy lies at the dynamical gap the off-diagonal terms together with higher order corrections must give a contribution of the same order as the one given by aforementioned theory to reproduce the exact numerical results.

\section{Hall response with ultra-short pulses}
In this section we briefly analyze the effect of ultra-short pulses on the Hall response and argue that it is possible to observe after-pulse topological Hall currents. Experiments able to measure ultrafast driven currents in clean graphene\cite{Obraztsov2014,Higuchi2017} have recently appeared, motivated by the fact that the control and optical manipulation of photocurrents in unbiased two dimensional samples might open new alternatives for photonics and optoelectronics. 
In fact, in Ref.~[\onlinecite{Higuchi2017}] after-pulse currents were measured,  showing that the carriers' lifetime is long enough to allow for a good characterization of the electron dynamics in time scales of the order of a few femtoseconds.

\begin{figure}[t]
\includegraphics[width=0.9\columnwidth]{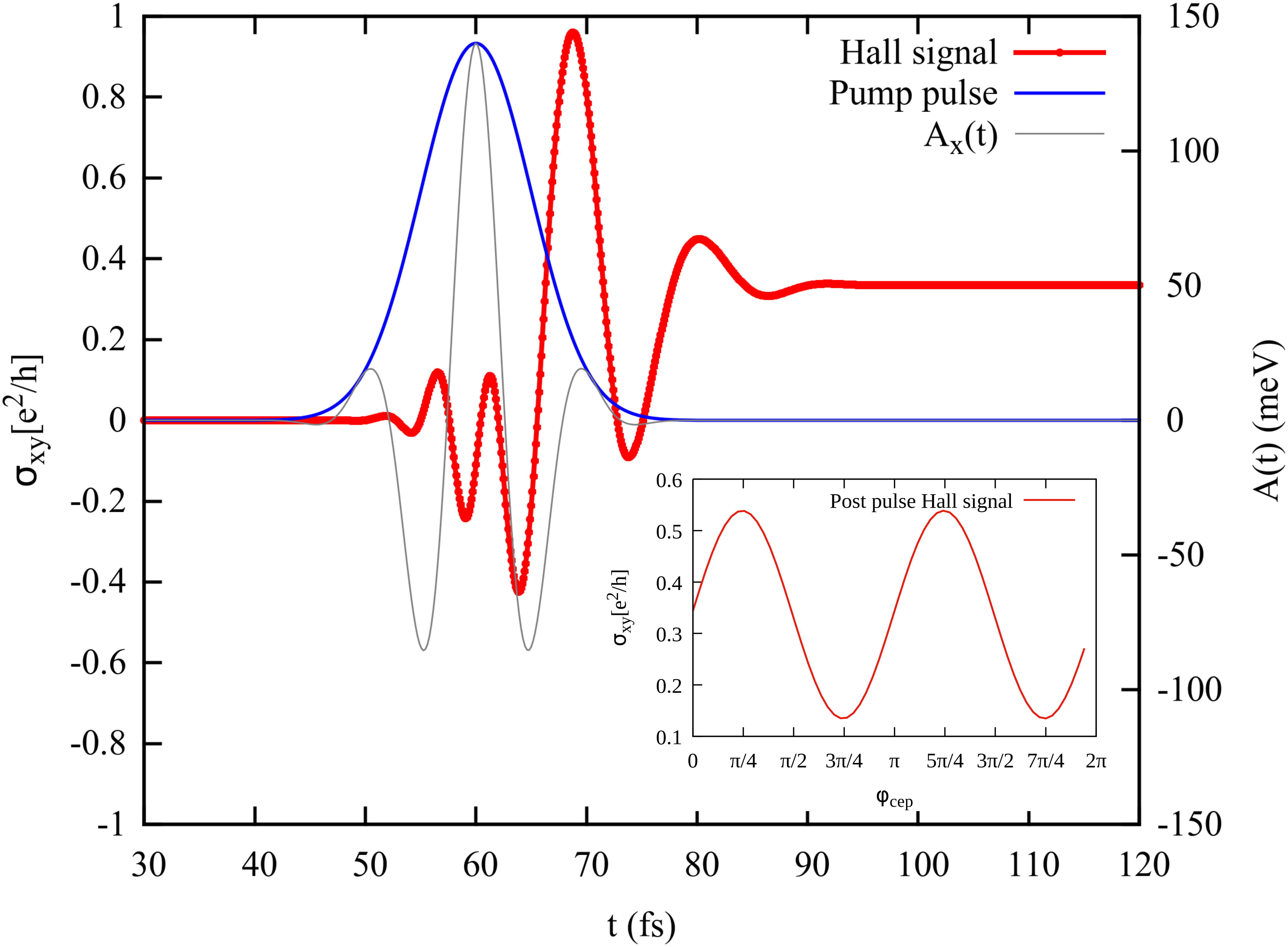}
\caption{Time-resolved Hall signal per valley with a pump envelope containing only a few cycles of the electromagnetic field. The dependence of the remanent post-pulse response with the carrier-envelope phase ($\varphi_{\text{CEP}}$) is shown in the inset.}
\label{fig9}
\end{figure}

Fig.~\ref{fig9} shows the Hall conductance as obtained with a Gaussian pulse containing only a few (between two and three) cycles of the carrier in the  undoped case. Within the pulse the $2\Omega$ oscillations are clearly observed, unveiling that even for these short perturbations the Floquet picture with the opening of gaps in the spectrum and the selection rule for the high frequency response give a good qualitative description.  For short pulses containing only a few cycles of the electromagnetic field, however, the system response depends on the carrier-envelope phase (CEP) $\varphi_{\text{CEP}}$--defined as the phase of the carrier measured from the maximum of the Gaussian envelope.   
This is shown in the inset of the figure, where we plot the asymptotic after-pulse response as a function of $\varphi_{\text{CEP}}$. The value of the mean after pulse Hall conductance, averaged on $\varphi_{\text{CEP}}$, is comparable to the ones obtained with wider pulses  in the previous section.
        
It is important to remark that even in the absence of the bias, small after-pulse currents can flow along the $x$ and $y$ directions~\cite{Higuchi2017} due to the non-zero time averaged electric field during the short pulse. However, in such a case the $\varphi_{\text{CEP}}$ averaged value is zero. With a bias field the after-pulse Hall response has a topological origin and its average is non-zero as can be inferred by the finite mean value of $\sigma_{xy}$ as a function of $\varphi_{\text{CEP}}$.

\section{Summary and Conclusions}

We have analyzed the full time-resolved Hall response of two-dimensional honeycomb lattices under coherent dynamics. We concentrate on systems like graphene and transition metal dichalcogenides. The results, however, are also relevant for a variety of systems showing dynamical topological properties. 
A simple toy model that switches a valley Hall signal shows how a dynamical response builds up in time when introducing a parameter ramp that breaks inversion symmetry in the lattice, with its asymptotic time averaged value depending on the particular ramping protocol. We have characterized the frequency and amplitude of the oscillating response and its dependence on doping.

Our central results concern the Hall response of these systems during and after short pulses of circularly polarized light and frequencies $\Omega$ smaller than their bandwidth. We have shown that in graphene like-systems, the Hall response develops a high-frequency signal with twice the driving frequency. The existence of this $2\Omega$ mode is a particular case of the heterodyne effect and the selection rule that supress other $\Omega$-multiples is due to the symmetry of the perturbed Hamiltonian.
After filtering these fast oscillations, the signal shows low frequency oscillations during the pulse. We can trace this response as coming from several Floquet gaps, unveiling the fact that some features of the photon-dressed topological bands are present even in a few femtosecond timescale. In fact, by a partial integration in $\bm{k}$-space it is possible to separate the contribution due to different regions of the BZ and consequently get information on their contribution.  In particular, each one of the gaps at the $\bm{K}$ and $\bm{K^\prime}$ points of the BZ has a low-frequency dependence that can be described by an effective slow-time evolution equation which is qualitative and quantitative similar to the switching of a mass term with different signs at each valley. This reflects the fact that the Floquet Hamiltonian mimics the occurrence of a dynamically achieved mass for each Dirac cone. Shifting the chemical potential to the dynamical gap at $-\hbar \Omega/2$, the main contribution to the Hall conductivity comes from states resonant with the photon energy. The low frequency oscillations are in good agreement with the Floquet gap  calculated at the lowest order, which is linear with the amplitude of the radiation field and independent of the driving frequency. 

Part of our analysis concerns the after pulse response. In graphene, all anomalous velocities of the after-pulse Hamiltonian, associated with the Berry curvature, are zero. Nonetheless, during the TRS breaking perturbation, the system undergoes a topological transition and the bias field induces anomalous velocities generating a transverse current. This current persists even after the pulse and is independent of the after-pulse bias field. In fact, if after the perturbation the bias is turned off, the current remains unaltered. In real systems these currents will of course decay with the characteristic scattering time of the electrons. However the long life-time of carriers in clean graphene allows measuring induced currents after ultra-short pulses~\citep{Higuchi2017}. Our results show that with ultra-short pulses, comprising only a few oscillations of the electromagnetic field, the after-pulse Hall currents could be detected.

We acknowledge financial support from PICTs 2013-1045 and 2016-0791 from ANPCyT, PIP 11220150100506 from CONICET and grant 06/C526 from SeCyT-UNC. 

\appendix
\section{Parameter quench on a two band model.\label{A}}

Consider a sudden perturbation that at time $t_0$ changes the Hamiltonian from its initial form ${\cal {H}}_{i}$ to a final form ${\cal {H}}_{f}$. For $t>t_0$ the wavefunctions and the corresponding time evolution operators are
\begin{eqnarray}
\ket{\psi_{\bm{k}\alpha}(t)}&=&\sum_{f}\Lambda_{f \alpha}^{\bm{k}}e^{-\frac{i}{\hbar}\varepsilon_{\bm{k}f}(t-t_0)}\ket{\phi_{\bm{k}f}}\\
U_{\bm{k}}(t,t^{\prime})&=&\sum_f e^{-\frac{i}{\hbar}\varepsilon_{\bm{k}f}(t-t^{\prime})}\ket{\phi_{\bm{k}f}}\bra{\phi_{\bm{k}f}}
\label{Us}
\end{eqnarray}
where $\varepsilon_{\bm{k}f}$ and $\ket{\phi_{\bm{k}f}}$ are the eigenvalues and eigenfunctions of ${\cal {H}}_{f}$ and $\Lambda_{f \alpha}^{\bm{k}}=\langle{\phi_{\bm{k}f}}\ket{\psi_{\bm{k}\alpha}(t_0)}$ is the projection of the final state $f$ on the eigenstate $\alpha$ of the unperturbed Hamiltonian. Using the above expressions in Eq.~(\ref{eq3}), we can calculate the mean Hall conductivity $\overline{\sigma}_{xy}$ by retaining only the diagonal ensamble of the final Hamiltonian, that is to say, dropping the off-diagonal terms of the density matrix. This  is equivalent to formally consider decoherence or cooling effects to capture the long-time behavior. Nevertheless, these off-diagonal components, ultimately arrising from a non-adiabatical evolution, unveil a rich behavior which we will address later on the discussion. 

In the particular case of a two band model, the contribution to the mean value can be expressed as
\begin{widetext}
\bea
\nonumber
\overline{\sigma}_{xy}(t) &=& \frac{e^2}{\hbar}\sum_{\bm{k}\alpha}\sum_f f(\varepsilon_{\bm{k}\alpha})|\Lambda_{f\alpha}^{\bm{k}}|^2\Bigg\{\Big[1-\cos\big(2\omega_{\bm{k}f}(t-t_0)\big) + 2\omega_{\bm{k}f}t_0\sin\big(2\omega_{\bm{k}f}(t-t_0)\big)\Big]\mathcal{F}_{xy}^{f}(\bm{k})\\
\label{A3}
&+&\Big[2\omega_{\bm{k}f}t - \sin\big(2\omega_{\bm{k}f}(t-t_0)\big)-2\omega_{\bm{k}f}t_0\cos\big(2\omega_{\bm{k}f}(t-t_0)\big)\Big]2\mathcal{G}_{xy}^{f}(\bm{k})\Bigg\},
\eea
\end{widetext}
with $\omega_{\bm{k}f} = \varepsilon_{\bm{k}f}/\hbar$. For $t-t_0 \rightarrow \infty$ we finally obtain

\bea
\label{A4}
\overline{\sigma}_{xy}(\infty) &=& \frac{e^2}{\hbar}\!\sum_{\bm{k}\alpha}\sum_f f(\varepsilon_{\bm{k}\alpha})|\Lambda^{\bm{k}}_{f\alpha}|^2\mathcal{F}_{xy}^f(\bm{k})\\
\nonumber
&=& \frac{e^2}{h}\frac{1}{2\pi}\sum_{\alpha}\int_{BZ}f(\varepsilon_{\bm{k}\alpha})\Big(|\Lambda^{\bm{k}}_{v\alpha}|^2 - |\Lambda^{\bm{k}}_{c\alpha}|^2\Big)\mathcal{F}_{xy}^{v}(\bm{k}),
\eea

where $v$ and $c$ stand for the valence and conduction band of the final Hamiltonian and we have used that $\mathcal{F}_{xy}^{c}(\bm{k}) = -\mathcal{F}_{xy}^{v}(\bm{k})$. The long-time limit of the mean Hall conductivity after a sudden parameter change is given by an integral of the Berry curvature of the final (time independent) Hamiltonian weighted by the occupation numbers of the after-quench states, a result already obtained by Wang et al~\cite{Wang2016}. The expression defined by Eq. (\ref{A4}) is a first indication that $\overline{\sigma}_{xy}(\infty)$ is not quantized and depends on the way the perturbation is turned on. It is important to keep in mind that the naively defined Chern number of the unitary evolved wave function
\begin{equation}
C(t) = \frac{1}{2\pi}\sum_{\alpha\,\epsilon\,\text{occ}}\int_{BZ}\mathcal{F}^{\alpha}_{xy}(\bm{k},t)d^2\bm{k}
\label{chern}
\end{equation}
with
\begin{equation}
\mathcal{F}^{\alpha}_{xy}(\bm{k},t) =\\
 -i\bigg[ \bigg\langle{\frac{\partial \psi_{\bm{k}\alpha}(t)}{\partial k_x}\bigg|\frac{\partial \psi_{\bm{k}\alpha}(t)}{\partial k_y}}\bigg\rangle -h.c. \bigg] 
\end{equation}
and $\ket{\psi_{\bm{k}\alpha}(t)}=U(t,-\infty)\ket{{\bm{k}\alpha}}$, does not manifest itself in the out-of-equilibrium Hall response. If this were the case, the topological transition could not be reflected in the Hall conductivity due to the preservation of $C(t)$ under a unitary evolution\cite{DAlessio2015}. 

In the particular case presented in Eq.~(\ref{Hmass}) we can calculate the response for a sudden switch of a mass-like term $\Delta(t)=\Delta_{\infty}\theta(t-t_0)$. Making use of the specific form of the Berry curvature and populations after the quench in Eq.~(\ref{A4}) we obtain for each valley at zero temperature 
\begin{equation}
\overline{\sigma}^\xi_{xy}(\!\infty\!)\!=\!-\xi\frac{e^2}{h}\!\int_{0}^{\infty}\!\frac{(\hbar v_f)^3 k^2 \Delta_{\infty}dk}{2[(\hbar v_f k)^2\!+\! \Delta_{\infty}^2]^2}\!=\!-\xi\frac{e^2}{h}\!\frac{\pi}{8}\text{sgn}(\!\Delta_{\infty}\!).
\label{A7}
\end{equation}
Even if this result isn't quantized with the topological invariant of the post-quench Hamiltonian, it is universal at zero temperature, in the sense of being independent of the magnitude of $\Delta_{\infty}$. Only within the hypothesis of a total adiabatic evolution, that can be easily obtained from Eq.~(\ref{A4}) by taking the population difference equal to unity, it is possible to recover the quantized value of the Hall response as the Chern of the final Hamiltonian. This limit is not expected to be accurate for ultra-short pulses or if the perturbation induces or eliminates degenerations in the energy spectrum. Nevertheless, the adiabatic presumption remains well suited for the evolution of Bloch states away from these situations. 

For finite times, closer to the initial ramp, a dynamical response is expected to be observed due to the oscillatory behavior with frequency $2\omega_{\bm{k}f}$ in the \textit{kernel} of integral Eq. (\ref{A3}). The observed oscillation after performing the integration will be governed by the energy gap at the quasi-momentums where the Berry curvature peaks. On the other hand, the calculation with the exact time evolution shows an increasing amplitude of the oscillations, which is not captured by the diagonal ensamble used to calculate the contribution to the mean Hall response. Their origin is then attributed to non-diagonal components of the density matrix in the basis of the final Hamiltonian, which we initially neglected.

We agglomerate this terms in a non-adiabatical contribution to the Hall conductivity $\sigma_{xy}^{NA}$, which can be expressed as
\begin{widetext}
\bea
\nonumber
\sigma_{xy}^{NA}(t)\!&=&\!\frac{e^2}{\hbar}\sum_{\bm{k}\alpha}\!\sum_{f\neq f'}\!f(\varepsilon_{\bm{k}\alpha})2\Re\Bigg\{\Lambda^{\bm{k}*}_{f'\alpha}\Lambda^{\bm{k}}_{f\alpha}\Bigg[\Bigg(\frac{1-e^{i\omega_{\bm{k}}^{f'f}(t-t_0)}}{\omega_{\bm{k}}^{f'f}}-i(t_0 - te^{i\omega_{\bm{k}}^{f'f}(t-t_0)})\Bigg)v_{x}^{\bm{k}f'}\!\mathcal{A}_{k_y}^{f'f}
+e^{-i\omega_{\bm{k}}^{ff'}(t-t_0)}\omega_{\bm{k}}^{ff'}\frac{t^2-t_0^2}{2}v_y^{\bm{k}f}\mathcal{A}_{k_x}^{f'f}\!\Bigg]\Bigg\},
\label{eqA8}\\
\!
\eea
\end{widetext}
where we defined the frequency $\omega_{\bm{k}}^{f'f} = (\varepsilon_{\bm{k}f'} - \varepsilon_{\bm{k}f})/\hbar$, the final states equilibrium velocity $v_{\nu}^{\bm{k}f} = \frac{1}{\hbar} \frac{\partial\varepsilon_{\bm{k}f}}{\partial k_{\nu}}$ and the Berry connection between two different states as $\mathcal{A}_{k_{\nu}}^{f'f} = -i\langle \phi_{\bm{k}f'}|\partial_{k_{\nu}}\phi_{\bm{k}f}\rangle$. The last term in Eq. (\ref{eqA8}), when performing the integral over the BZ, will retain an oscillating character but with amplitudes increasing in time as $\sim t + t_0$. 

\section{Heterodyne effect.}
It is well known that when driving a system with periodic frequency $\Omega$ and measuring a response function generated by an input signal (in this case the bias current along the $\hat{y}$ direction), it oscillates with multiples of the driving frequency~\cite{Oka2016}. In our time-resolved Hall conductivity simulations a clear selection rule brings out a response with a manifesting mode at $2\Omega$, besides the small frequency oscillations associated with the Floquet gaps. 

Indeed, this selection rule can be traced to the angular integration of the correlation functions in momentum space. We define the unitary transformed state vector as
\begin{equation}
\mathbf{\hat{R}}(\theta_{\bm{k}})|\psi_{\bm{k}}^{\alpha}(t,\tau)\rangle\rangle = |\widetilde{\psi}_{\bm{k}}^{\alpha}(t,\tau)\rangle\rangle,
\end{equation}
with $\mathbf{\hat{R}}(\theta_{\bm{k}}) = e^{i\frac{\theta_{\bm{k}}}{2}\sigma_z}$ and $\theta_{\bm{k}} = \text{tan}^{-1}(k_y/k_x)$. This rotation is performed in order to leave the initial unperturbed Dirac spinors at the $\hat{x}$ axes. The transformed time dependent wavefunction verifies an extended Schr\"odinger equation of the form
\begin{equation}
[\widetilde{H}_{\bm{k}}(t,\tau) - i\hbar\partial_t]|\widetilde{\psi}_{\bm{k}}^{\alpha}(t,\tau)\rangle\rangle = i\hbar\partial_{\tau}|\widetilde{\psi}_{\bm{k}}^{\alpha}(t,\tau)\rangle\rangle
\label{eq4}
\end{equation}
with
\begin{eqnarray}
\widetilde{H}_{\bm{k}}(t,\tau)\!&=&\!\mathbf{\hat{R}}(\theta_{\bm{k}})H_{\bm{k}}(t,\tau)\mathbf{\hat{R}^{\dagger}}(\theta_{\bm{k}})\\
\notag
\!&=&\!\hbar v_f k \sigma_x\!+\!e v_f A(\!\tau\!)\![\cos(\Omega t\!-\!\theta_{\bm{k}})\sigma_x\!+\!\sin(\Omega t\!-\! \theta_{\bm{k}})\sigma_y\!].
\end{eqnarray} 

We extend this transformed state vector in its Fourier modes by taking into account the angle $\theta_{\bm{k}}$ as an initial $\bm{k}$-dependent phase, \textit{i.e.} $\Omega t_0(\bm{k}) = \theta_{\bm{k}}$,
\begin{equation}
|\widetilde{\psi}^{\alpha}_{\bm{k}}(t,\tau)\rangle\rangle = \sum_{n}e^{in(\Omega t -\theta_{\bm{k}})}\widetilde{\chi}_{n\bm{k}}(\tau).
\label{eq6}
\end{equation}

The inital condition $\widetilde{\chi}_{n\bm{k}}(-\infty) = \hat{R}(\theta_{\bm{k}})|\psi_{\bm{k}\alpha}\rangle\delta_{n,0} = |\pm\rangle_{x}\delta_{n,0}$ is such that the Bloch states in this basis are originally eigenstates of $\sigma_x$, independent of $\bm{k}$. Replacing Eq. (\ref{eq6}) into Eq. (\ref{eq4}) we get an infinitly coupled set of equations
\begin{widetext}
\begin{equation}
(\hbar v_f k \sigma_x + n\hbar\Omega)\widetilde{\chi}_{n\bm{k}}(\tau) + V(\tau)\widetilde{\chi}_{n-1,\bm{k}}(\tau) + V^{\dagger}(\tau)\widetilde{\chi}_{n+1,\bm{k}}(\tau)= i\hbar\partial_{\tau}\widetilde{\chi}_{n,\bm{k}}(\tau),
\label{eq7}
\end{equation}
\end{widetext}
with $V(\tau) = e v_f A(\tau)\frac{\sigma_x - i\sigma_y}{2}$. 
The initial condition and the slow time evolution determined by Eq. (\ref{eq7}) are independent of the angle $\theta_{\bm{k}}$, so in this basis the modes $\widetilde{\chi}_{n\bm{k}}(\tau)$ are angle-independent at all times. This simple procedure shows that the only dependence of Floquet modes with this angle is exponential with a factor $n$. When integrating in polar coordinates near the Dirac cones a simple selection rule is found. The Hall conductivity can be written as

\begin{widetext}
\begin{eqnarray}
\label{eqB6}
\sigma_{xy}(t)&=&\frac{e^2}{h}\frac{v_f^2}{\pi}\Im\Big[\!\sum_{\alpha,\beta}\!\sum_{\substack{n,n'\\l,l'}}f(\varepsilon_{\bm{k}\alpha})\int_{-\infty}^{t}\!dt'\int_{0}^{\infty}\! k dk\int_{0}^{2\pi}\! d\theta_{\bm{k}}e^{i(n-n'+l-l')\theta_{\bm{k}}}e^{i(n-n')\Omega t}e^{i(l-l')\Omega t'}\!\times\\
\notag
& &\langle\widetilde{\chi}^{\alpha}_{n'k}(\tau)|\sigma_x(\theta_{\bm{k}})|\widetilde{\chi}^{\beta}_{nk}(\tau)\rangle\langle\widetilde{\chi}^{\beta}_{l'k}(\tau')|\sigma_y(\theta_{\bm{k}})|\widetilde{\chi}^{\alpha}_{lk}(\tau')\rangle\mathcal{W}(t')\Big]
\end{eqnarray}
\end{widetext}
where the rotated Pauli matrices are

\begin{eqnarray}
\sigma_x(\theta_{\bm{k}}) &=& \cos(\theta_{\bm{k}})\sigma_x - \sin(\theta_{\bm{k}})\sigma_y\\
\notag
\sigma_y(\theta_{\bm{k}}) &=& \sin(\theta_{\bm{k}})\sigma_x + \cos(\theta_{\bm{k}})\sigma_y.
\end{eqnarray}

If we perform the angular integral in Eq. (\ref{eqB6}) we get that $n-n'+l-l' = 0,\pm 2$, while all the other contributions cancel out. When  $n-n'+l-l' = 0$ the only time dependent oscillations will come from the slow dynamics, since the \textit{kernel} of the time-integral depends on $e^{i(l-l')\Omega(t'-t)}$ and hence the result will not retain the fast oscillations. On the other hand, when $n-n'+l-l' = \pm 2$ the contribution remains oscillatory with frequency $2\Omega$.

\section{Slow time dynamics near the Dirac points.}
The appearance of a gap in the quasi-energy Floquet spectrum at the Dirac points is due to a virtual process of absortion and emission of a photon, therefore its magnitude shows a quadratic dependence with the field strength. We search for a canonical transformation of the Floquet operator $H_F^{\infty}(\xi,\tau)$ that retains quadratic orders in the radiation field in order to obtain a dynamical effective equation that describes accurately the slow time evolution near these gaps. We look for a unitary transformation where the transformed state vector
\begin{equation}
\widetilde{\chi}_{\xi}(\tau)= e^{-S_{\xi}(\tau)}\bar{\chi}_{\xi}(\tau),
\label{eqc}
\end{equation}
obeys a Floquet-Schr\"odinger time dependent equation (see Eq. (\ref{floquetscrho}) in the main text) with a modified Floquet-operator of the form
\begin{eqnarray}
\label{eqa}
\tilde{H}_F^{\infty}(\xi,\tau)\widetilde{\chi}_{\xi}(\tau)=i\hbar\partial_{\tau}\widetilde{\chi}_{\xi}(\tau),\,\,\,\,\,\,\,\,\,\,\,\,\,\,\,\,\,\,\,\,\,\,\,\,\,\,\\
\notag
\tilde{H}_F^{\infty}(\xi,\tau)\!=\!e^{-S_{\xi}(\tau)}\! H_F^{\infty}(\xi,\tau)e^{S_{\xi}(\tau)}\!-\! i\hbar e^{-S_{\xi}(\tau)}\!\frac{d}{d\tau}\!e^{S_{\xi}(\tau)}.
\end{eqnarray}
Notice that due to the additional temporal dependence of the unitary transformation there is an additional term that involves time derivatives of $S_{\xi}(\tau)$. Separating the unperturbed block-diagonal part of the hamiltonian $H^{\infty}_0(\xi)$ from the time dependent coupling we have that $H_F^{\infty}(\xi,\tau) = H^{\infty}_0(\xi) + e v_f A(\tau)H'^{\infty}_{\xi}$, where
\begin{equation}
H^{\infty}_0(\xi)\!=\!\left(\begin{array}{ccccc}
\ddots\!&\!\vdots\!&\!\vdots\!&\!\vdots\!&\!\reflectbox{$\ddots$}\\
\ldots\!&\!H^{0}_{\xi}(\bm{k})\!+\!\hbar\Omega\bm{I}\!&\!\bm{0}\!&\!\bm{0}\!&\!\ldots\\
\ldots\!&\!\bm{0}\!&\!H^{0}_{\xi}(\bm{k})\!&\!\bm{0}\!&\!\ldots\\
\ldots\!&\!\bm{0}\!&\!\bm{0}\!&\!H^{0}_{\xi}(\bm{k})\!-\!\hbar\Omega\bm{I} \!&\!\ldots\\
\reflectbox{$\ddots$}\!&\!\vdots\!&\!\vdots\!&\!\vdots\!&\!\ddots\\
\end{array} \right)
\end{equation}
and
\begin{equation}
e v_f A(\tau) H'^{\infty}_{\xi}\!=\!\left(\!\begin{array}{ccccc}
\ddots\!&\!\vdots\!&\!\vdots\!&\!\vdots\!&\!\reflectbox{$\ddots$}\\
\ldots\!&\!\bm{0}\!&\!H^{(1)}_{\xi}(\tau)\!&\!\bm{0}\!&\! \ldots\\
\ldots\!&\!H^{(-1)}_{\xi}(\tau)\!&\!\bm{0}\!&\!H^{(1)}_{\xi}(\tau)\!&\!\ldots\\
\ldots\!&\!\bm{0}\!&\!H^{(-1)}_{\xi}(\tau)\!&\!\bm{0} &\ldots\\
\reflectbox{$\ddots$}\!&\!\vdots\!&\!\vdots\!&\!\vdots\!&\!\ddots\\
\end{array}\!\right).
\end{equation}
Here we have specificed the case of the Dirac hamiltonian driven with a circularly polarized laser, where the Fourier components involved are $H^{0}_{\xi}(\bm{k}) = \hbar v_f (\xi k_x\sigma_x + k_y\sigma_y)$ and $H^{(\pm 1)}_{\xi}(\tau) = e v_f A(\tau)\frac{\xi\sigma_x \mp i\sigma_y}{2}$ for each valley. 

Expanding Eq. (\ref{eqa}) up to second order in $S_{\xi}(\tau)$ it's easy to see that linear terms in $A(\tau)$ will vanish if $S_{\xi}(\tau)$ is chosen to make
\begin{equation}
e v_f A(\tau)H'^{\infty}(\xi) + [H^{\infty}_0(\xi),S_{\xi}(\tau)] = 0.
\end{equation}
This last identity verifies near the Dirac points taking the canonical transformation to be
\begin{equation}
S_{\xi}(\tau)\!=\!\left(\begin{array}{ccccc}
\ddots\!&\!\vdots\!&\!\vdots\!&\!\vdots\!&\!\reflectbox{$\ddots$}  \\
\ldots\!&\!\bm{0}\!&\!-\!\eta(\tau)\sigma_{-}(\xi)\!&\!\bm{0}\!&\!\ldots\\
\ldots\!&\!\eta(\tau)\sigma_{+}(\xi)\!&\!\bm{0}\!&\!-\eta(\tau)\sigma_{-}(\xi)\!&\!\ldots\\
\ldots\!&\!\bm{0}\!&\!\eta(\tau)\sigma_{+}(\xi)\!&\!\bm{0}\!&\!\ldots\\
\reflectbox{$\ddots$}\!&\!\vdots\!&\!\vdots\!&\!\vdots\!&\!\ddots\\
\end{array} \right),
\end{equation}
with $\eta(\tau) = \frac{e v_f A(\tau)}{\hbar\Omega}$ and $\sigma_{\pm}(\xi) = \frac{\xi\sigma_x \pm i\sigma_y}{2}$. We then obtain an effective dynamical Floquet-Schr\"odinger equation to order $A^2(\tau)$ for the state-vector $\widetilde{\chi}_{\xi}(\tau)$
\begin{widetext}
\begin{equation}
\begin{split}
\Big[H^{\infty}_0(\xi) + \frac{e v_f A(\tau)}{2c}[H'^{\infty}_{\xi},S_{\xi}(\tau)] - i\hbar\dot{S}_{\xi}(\tau) + \mathcal{O}[A^3(\tau)]\Big]\widetilde{\chi}_{\xi}(\tau) = i\hbar\partial_{\tau}\widetilde{\chi}_{\xi}(\tau),\,\,\,\,\,\,\,\,\,\,\,\,\,\,\,\,\,\,\,\,\,\,\\
{\left( \begin{array}{ccccc}
\ddots & \vdots & \vdots &\vdots &\reflectbox{$\ddots$}  \\
\ldots & H_{\text{eff}}(\xi,\tau) + \hbar\Omega\bm{I} & i\hbar\dot{\eta}(\tau)\sigma_{-}(\xi) & \bm{0} & \ldots\\
\ldots & -i\hbar\dot{\eta}(\tau)\sigma_{+}(\xi) & H_{\text{eff}}(\xi,\tau) & i\hbar\dot{\eta}(\tau)\sigma_{-}(\xi) &\ldots\\
\ldots & \bm{0} & -i\hbar\dot{\eta}(\tau)\sigma_{+}(\xi) & H_{\text{eff}}(\xi,\tau) - \hbar\Omega\bm{I} &\ldots\\
\reflectbox{$\ddots$} & \vdots & \vdots & \vdots & \ddots\\
\end{array} \right)\left(\begin{array}{c}
\vdots \\
\widetilde{\chi}_1(\xi,\tau)\\
\widetilde{\chi}_0(\xi,\tau)\\
\widetilde{\chi}_{-1}(\xi,\tau)\\
\vdots
\end{array}\right) = i\hbar\partial_{\tau}\left(\begin{array}{c}
\vdots \\
\widetilde{\chi}_1(\xi,\tau)\\
\widetilde{\chi}_0(\xi,\tau)\\
\widetilde{\chi}_{-1}(\xi,\tau)\\
\vdots
\end{array}\right)}
\end{split}
\label{eqb}
\end{equation}
\end{widetext}

where we used that $[S_{\xi}(\tau),\dot{S}_{\xi}(\tau)] = 0$ and defined an effective hamiltonian
\begin{eqnarray}
\notag
H_{\text{eff}}(\xi,\tau)&=&H_{\xi}^{0}(\bm{k}) + \frac{1}{\hbar\Omega}[H^{(1)}_{\xi}(\tau),H^{(-1)}_{\xi}(\tau)]\\
&=& H_{\xi}^{0}(\bm{k}) - \frac{[e v_f A(\tau)]^2}{\hbar\Omega}\xi\sigma_z.
\end{eqnarray}

If we take the limit $\dot{\eta}(\tau) = \frac{e v_f \dot{A}(\tau)}{\hbar\Omega}\rightarrow 0$ the diagonal blocks are uncoupled. It's important to notice that this approximation is less restrictive than the total adiabatic limit. Even if the derivative of the pulse envelope remains finite, this limit would be appropriate for sufficiently high photon energies, which is the case of interest in the non-resonant regime. The time dependent equations for the different modes $\widetilde{\chi}_n(\xi,\tau)$ can be then written as
\begin{eqnarray}
[\!H_{\text{eff}}(\xi,\!\tau\!)\!+\! n\hbar\Omega\bm{I}_2\!]\widetilde{\chi}_n(\xi,\!\tau\!)\!&=&\! i\hbar\partial_{\tau}\widetilde{\chi}_n(\xi,\!\tau\!)\\
\notag
\widetilde{\chi}_n(\xi,\!\tau\!)\!&=&\!e^{-in\Omega \tau}U^{\xi}_{\text{eff}}(\tau,\!\tau_0\!)\widetilde{\chi}_n(\xi,\tau_0),
\label{eq20}
\end{eqnarray}
where $\bm{I}_2$ is a $2\text{x}2$ identity matrix and the effective evolution operator is formally obtained as
\begin{equation}
U^{\xi}_{\text{eff}}(\tau,\tau_0)=\mathcal{T}\Big[e^{-\frac{i}{\hbar}\int_{\tau_0}^{\tau}H_{\text{eff}}(\xi,\tau')d\tau'}\Big],
\end{equation}
with $\mathcal{T}$ the time ordering operator. The initial condition is chosen to have only the zero mode occupied and in the valence band, \textit{i.e.} $\chi_n(\xi,\tau=\tau_0) = \delta_{n,0}\ket{\psi_{\bm{k}v \xi}}$. Since $\eta(\tau_0) = 0$, the unitary transformation in Eq. (\ref{eqc}) reduces to the identity matrix, making $\widetilde{\chi}_n(\xi,\tau=\tau_0) = \chi_n(\xi,\tau=\tau_0)$.

Identifying $\Delta(\xi,\tau) = -\xi\frac{[e v_f A(\tau)]^2}{c^2\hbar\Omega}$, it is understood that the slow time dynamics of any observable quantity near these valleys will be determined by a Dirac hamiltonian with a switching mass-like term $\Delta(\xi,\tau)$ proportional to $\sigma_z$. Finally, in the interest of expressing the statevector in its original basis, we simply apply the inverse transformation in Eq. (\ref{eqc}) to second order in $S_{\xi}(\tau)$ obtaining an effective evolution for the modes $n = 0,\pm 1$ given by
\begin{eqnarray}
\notag
\chi_0(\xi,\tau) &=&\mathcal{I}^{2\times 2}\Big(1-\frac{\eta^2(\tau)}{2}\Big)U^{\xi}_{\text{eff}}(\tau,\!\tau_0\!)\chi_0(\xi,\tau_0)\\
\notag
\chi_1(\xi,\tau) &=& -\eta(\tau)\sigma_{-}(\xi)U^{\xi}_{\text{eff}}(\tau,\!\tau_0\!)\chi_0(\xi,\tau_0)\\
\chi_{-1}(\xi,\tau) &=& \eta(\tau)\sigma_{+}(\xi)U^{\xi}_{\text{eff}}(\tau,\!\tau_0\!)\chi_0(\xi,\tau_0).
\end{eqnarray}

Rewriting the extended two-time wavefunction as $\Ket{\psi_{\xi}(\tau,t)} = \sum_n e^{in\Omega t}\chi_n(\xi,\tau)$ and restricting the solution to the physical contour $\tau = t$ we obtain for the approximated wavefunction
\begin{widetext}
\begin{equation}
|\psi_{\xi}(\tau,t)\rangle\rangle\Big\rvert_{\tau = t} = |\psi_{\xi}(t)\rangle = \Big[\mathcal{I}^{2\times 2}\Big(1-\frac{\eta^2(\tau)}{2}\Big) + \eta(\tau)[\sigma_{+}(\xi)e^{-i\Omega\tau}-\sigma_{-}(\xi)e^{i\Omega\tau}]\Big]U^{\xi}_{\text{eff}}(\tau,\!\tau_0\!)\widetilde{\chi}_0(\xi,\tau_0),
\label{appwf}
\end{equation}
\end{widetext}
which is normalized to order $\eta^2(t)$. This procedure guarantees an effective evolution that allows tunneling between Floquet modes  during the pulse duration.

If we use Eq. (\ref{appwf}) to calculate the contribution to the Hall conductance near these valleys we get

\begin{equation}
\sigma_{xy}(t) = \frac{e^2 v_f^2}{i\hbar}\sum_{\bm{k}\alpha}\int_{-\infty}^{t}\langle\psi_{\bm{k}\alpha}|\mathcal{C}_{\bm{k}}(t,t')|\psi_{\bm{k}\alpha}\rangle\mathcal{W}(t')dt',
\label{corr}
\end{equation}
with
\begin{widetext}
\begin{eqnarray}
\notag
C_{\bm{k}}(t,t') &=& \Big(1-\eta^2(t)-\eta^2(t')\Big)[\xi\sigma_x^{\text{eff}}(t),\sigma_y^{\text{eff}}(t')] - 2\xi\eta(t)\cos(\Omega t) [\sigma_z^{\text{eff}}(t),\sigma_y^{\text{eff}}(t')]\\
\notag
&-& 2\xi\eta(t')\sin(\Omega t') [\xi\sigma_x^{\text{eff}}(t),\sigma_z^{\text{eff}}(t')] + 4\eta(t)\eta(t')\cos(\Omega t)\sin(\Omega t')[\sigma_z^{\text{eff}}(t),\sigma_z^{\text{eff}}(t')]\\
\notag
&+&\eta^2(t')\Big(ie^{2i\Omega t'}[\xi\sigma_x^{\text{eff}}(t),\sigma_{-}^{\text{eff}}(\xi,t')] - ie^{-2i\Omega t'}[\xi\sigma_x^{\text{eff}}(t),\sigma_{+}^{\text{eff}}(\xi,t')]\Big)\\
&-&\eta^2(t)\Big(e^{2i\Omega t}[\sigma_{-}^{\text{eff}}(\xi,t),\sigma_y^{\text{eff}}(t')]+e^{-2i\Omega t}[\sigma_{+}^{\text{eff}}(\xi,t),\sigma_y^{\text{eff}}(t')]\Big),
\label{eq26}
\end{eqnarray}
\end{widetext}
where we use the notation $\mathcal{O}^{\text{eff}}(t) = U_{\text{eff}}(t_0,t)\mathcal{O}U_{\text{eff}}(t,t_0)$ for the operators written in the interaction picture with the effective evolution operator introducen in Eq. (\ref{eq20}). Using the same argument that in Appendix \textbf{B} we can show that the only oscillating factors surviving the angular integration are the ones with $2\Omega$.

%

\begin{thebibliography}{52}%
\makeatletter
\providecommand \@ifxundefined [1]{%
 \@ifx{#1\undefined}
}%
\providecommand \@ifnum [1]{%
 \ifnum #1\expandafter \@firstoftwo
 \else \expandafter \@secondoftwo
 \fi
}%
\providecommand \@ifx [1]{%
 \ifx #1\expandafter \@firstoftwo
 \else \expandafter \@secondoftwo
 \fi
}%
\providecommand \natexlab [1]{#1}%
\providecommand \enquote  [1]{``#1''}%
\providecommand \bibnamefont  [1]{#1}%
\providecommand \bibfnamefont [1]{#1}%
\providecommand \citenamefont [1]{#1}%
\providecommand \href@noop [0]{\@secondoftwo}%
\providecommand \href [0]{\begingroup \@sanitize@url \@href}%
\providecommand \@href[1]{\@@startlink{#1}\@@href}%
\providecommand \@@href[1]{\endgroup#1\@@endlink}%
\providecommand \@sanitize@url [0]{\catcode `\\12\catcode `\$12\catcode
  `\&12\catcode `\#12\catcode `\^12\catcode `\_12\catcode `\%12\relax}%
\providecommand \@@startlink[1]{}%
\providecommand \@@endlink[0]{}%
\providecommand \url  [0]{\begingroup\@sanitize@url \@url }%
\providecommand \@url [1]{\endgroup\@href {#1}{\urlprefix }}%
\providecommand \urlprefix  [0]{URL }%
\providecommand \Eprint [0]{\href }%
\providecommand \doibase [0]{http://dx.doi.org/}%
\providecommand \selectlanguage [0]{\@gobble}%
\providecommand \bibinfo  [0]{\@secondoftwo}%
\providecommand \bibfield  [0]{\@secondoftwo}%
\providecommand \translation [1]{[#1]}%
\providecommand \BibitemOpen [0]{}%
\providecommand \bibitemStop [0]{}%
\providecommand \bibitemNoStop [0]{.\EOS\space}%
\providecommand \EOS [0]{\spacefactor3000\relax}%
\providecommand \BibitemShut  [1]{\csname bibitem#1\endcsname}%
\let\auto@bib@innerbib\@empty
\bibitem [{\citenamefont {von Klitzing}\ \emph {et~al.}(1980)\citenamefont {von
  Klitzing}, \citenamefont {Dorda},\ and\ \citenamefont
  {Pepper}}]{Klitzing1980}%
  \BibitemOpen
  \bibfield  {author} {\bibinfo {author} {\bibfnamefont {K.}~\bibnamefont {von
  Klitzing}}, \bibinfo {author} {\bibfnamefont {G.}~\bibnamefont {Dorda}}, \
  and\ \bibinfo {author} {\bibfnamefont {M.}~\bibnamefont {Pepper}},\
  }\bibfield  {title} {\enquote {\bibinfo {title} {New method for high-accuracy
  determination of the fine-structure constant based on quantized {H}all
  resistance},}\ }\href {\doibase 10.1103/PhysRevLett.45.494} {\bibfield
  {journal} {\bibinfo  {journal} {Phys. Rev. Lett.}\ }\textbf {\bibinfo
  {volume} {45}},\ \bibinfo {pages} {494} (\bibinfo {year} {1980})}\BibitemShut
  {NoStop}%
\bibitem [{\citenamefont {Laughlin}(1981)}]{Laughlin1981}%
  \BibitemOpen
  \bibfield  {author} {\bibinfo {author} {\bibfnamefont {R.}~\bibnamefont
  {Laughlin}},\ }\bibfield  {title} {\enquote {\bibinfo {title} {Quantized
  {H}all conductivity in two dimensions},}\ }\href@noop {} {\bibfield
  {journal} {\bibinfo  {journal} {Phys. Rev. B}\ }\textbf {\bibinfo {volume}
  {23}},\ \bibinfo {pages} {5632} (\bibinfo {year} {1981})}\BibitemShut
  {NoStop}%
\bibitem [{\citenamefont {Halperin}(1982)}]{Halperin1982a}%
  \BibitemOpen
  \bibfield  {author} {\bibinfo {author} {\bibfnamefont {B.}~\bibnamefont
  {Halperin}},\ }\bibfield  {title} {\enquote {\bibinfo {title} {Quantized
  {H}all conductance, current-carrying edge states, and the existence of
  extended states in a two-dimensional disordered potential},}\ }\href@noop {}
  {\bibfield  {journal} {\bibinfo  {journal} {Phys. Rev. B}\ }\textbf {\bibinfo
  {volume} {25}},\ \bibinfo {pages} {2185} (\bibinfo {year}
  {1982})}\BibitemShut {NoStop}%
\bibitem [{\citenamefont {Thouless}\ \emph {et~al.}(1982)\citenamefont
  {Thouless}, \citenamefont {Kohmoto}, \citenamefont {Nightingale},\ and\
  \citenamefont {den Nijs}}]{Thouless1982}%
  \BibitemOpen
  \bibfield  {author} {\bibinfo {author} {\bibfnamefont {D.~J.}\ \bibnamefont
  {Thouless}}, \bibinfo {author} {\bibfnamefont {M.}~\bibnamefont {Kohmoto}},
  \bibinfo {author} {\bibfnamefont {M.~P.}\ \bibnamefont {Nightingale}}, \ and\
  \bibinfo {author} {\bibfnamefont {M.}~\bibnamefont {den Nijs}},\ }\bibfield
  {title} {\enquote {\bibinfo {title} {Quantized {H}all conductance in a
  two-dimensional periodic potential},}\ }\href {\doibase
  10.1103/PhysRevLett.49.405} {\bibfield  {journal} {\bibinfo  {journal} {Phys.
  Rev. Lett.}\ }\textbf {\bibinfo {volume} {49}},\ \bibinfo {pages} {405}
  (\bibinfo {year} {1982})}\BibitemShut {NoStop}%
\bibitem [{\citenamefont {Hasan}\ and\ \citenamefont {Kane}(2010)}]{Hasan2010}%
  \BibitemOpen
  \bibfield  {author} {\bibinfo {author} {\bibfnamefont {M.~Z.}\ \bibnamefont
  {Hasan}}\ and\ \bibinfo {author} {\bibfnamefont {C.~L.}\ \bibnamefont
  {Kane}},\ }\bibfield  {title} {\enquote {\bibinfo {title} {Colloquium:
  Topological insulators},}\ }\href {\doibase 10.1103/RevModPhys.82.3045}
  {\bibfield  {journal} {\bibinfo  {journal} {Rev. Mod. Phys.}\ }\textbf
  {\bibinfo {volume} {82}},\ \bibinfo {pages} {3045} (\bibinfo {year}
  {2010})}\BibitemShut {NoStop}%
\bibitem [{\citenamefont {Kane}\ and\ \citenamefont {Moore}(2011)}]{Kane2011}%
  \BibitemOpen
  \bibfield  {author} {\bibinfo {author} {\bibfnamefont {C.}~\bibnamefont
  {Kane}}\ and\ \bibinfo {author} {\bibfnamefont {J.}~\bibnamefont {Moore}},\
  }\bibfield  {title} {\enquote {\bibinfo {title} {Topological insulators},}\
  }\href@noop {} {\bibfield  {journal} {\bibinfo  {journal} {Physics World,
  february}\ }\textbf {\bibinfo {volume} {32}} (\bibinfo {year}
  {2011})}\BibitemShut {NoStop}%
\bibitem [{\citenamefont {Ando}(2013)}]{Ando2013}%
  \BibitemOpen
  \bibfield  {author} {\bibinfo {author} {\bibfnamefont {Y.}~\bibnamefont
  {Ando}},\ }\bibfield  {title} {\enquote {\bibinfo {title} {Topological
  insulator materials},}\ }\href@noop {} {\bibfield  {journal} {\bibinfo
  {journal} {J. Phys. Soc. Jpn.}\ }\textbf {\bibinfo {volume} {82}},\ \bibinfo
  {pages} {102001} (\bibinfo {year} {2013})}\BibitemShut {NoStop}%
\bibitem [{\citenamefont {Bernevig}\ and\ \citenamefont
  {Hughes}(2013)}]{Bernevig2013}%
  \BibitemOpen
  \bibfield  {author} {\bibinfo {author} {\bibfnamefont {B.~A.}\ \bibnamefont
  {Bernevig}}\ and\ \bibinfo {author} {\bibfnamefont {T.~L.}\ \bibnamefont
  {Hughes}},\ }\href@noop {} {\emph {\bibinfo {title} {Topological Insulators
  and Topological Superconductors}}}\ (\bibinfo  {publisher} {Princeton
  University Press},\ \bibinfo {year} {2013})\BibitemShut {NoStop}%
\bibitem [{\citenamefont {Shen}(2013)}]{Shen2013}%
  \BibitemOpen
  \bibfield  {author} {\bibinfo {author} {\bibfnamefont {S.-Q.}\ \bibnamefont
  {Shen}},\ }\href@noop {} {\emph {\bibinfo {title} {Topological Insulators:
  Dirac Equation in Condensed Matters}}},\ \bibinfo {edition} {2013th}\ ed.,\
  Springer Series in Solid-State Sciences (Book 174)\ (\bibinfo  {publisher}
  {Springer},\ \bibinfo {year} {2013})\BibitemShut {NoStop}%
\bibitem [{\citenamefont {Haldane}(1988)}]{Haldane1988}%
  \BibitemOpen
  \bibfield  {author} {\bibinfo {author} {\bibfnamefont {F.~D.~M.}\
  \bibnamefont {Haldane}},\ }\bibfield  {title} {\enquote {\bibinfo {title}
  {Model for a quantum {H}all effect without {L}andau levels: Condensed-matter
  realization of the ``parity anomaly"},}\ }\href {\doibase
  10.1103/PhysRevLett.61.2015} {\bibfield  {journal} {\bibinfo  {journal}
  {Phys. Rev. Lett.}\ }\textbf {\bibinfo {volume} {61}},\ \bibinfo {pages}
  {2015} (\bibinfo {year} {1988})}\BibitemShut {NoStop}%
\bibitem [{\citenamefont {Nagaosa}\ \emph {et~al.}(2010)\citenamefont
  {Nagaosa}, \citenamefont {Sinova}, \citenamefont {Onoda}, \citenamefont
  {MacDonald},\ and\ \citenamefont {Ong}}]{Nagaosa2010}%
  \BibitemOpen
  \bibfield  {author} {\bibinfo {author} {\bibfnamefont {N.}~\bibnamefont
  {Nagaosa}}, \bibinfo {author} {\bibfnamefont {J.}~\bibnamefont {Sinova}},
  \bibinfo {author} {\bibfnamefont {S.}~\bibnamefont {Onoda}}, \bibinfo
  {author} {\bibfnamefont {A.~H.}\ \bibnamefont {MacDonald}}, \ and\ \bibinfo
  {author} {\bibfnamefont {N.~P.}\ \bibnamefont {Ong}},\ }\bibfield  {title}
  {\enquote {\bibinfo {title} {Anomalous {H}all effect},}\ }\href@noop {}
  {\bibfield  {journal} {\bibinfo  {journal} {Reviews of Modern Physics}\
  }\textbf {\bibinfo {volume} {82}},\ \bibinfo {pages} {1539} (\bibinfo {year}
  {2010})}\BibitemShut {NoStop}%
\bibitem [{\citenamefont {Oka}\ and\ \citenamefont {Aoki}(2009)}]{Oka2009}%
  \BibitemOpen
  \bibfield  {author} {\bibinfo {author} {\bibfnamefont {T.}~\bibnamefont
  {Oka}}\ and\ \bibinfo {author} {\bibfnamefont {H.}~\bibnamefont {Aoki}},\
  }\bibfield  {title} {\enquote {\bibinfo {title} {Photovoltaic {H}all effect
  in graphene},}\ }\href@noop {} {\bibfield  {journal} {\bibinfo  {journal}
  {Phys. Rev. B}\ }\textbf {\bibinfo {volume} {79}},\ \bibinfo {pages} {081406}
  (\bibinfo {year} {2009})}\BibitemShut {NoStop}%
\bibitem [{\citenamefont {Oka}\ and\ \citenamefont {Aoki}(2010)}]{Oka2010}%
  \BibitemOpen
  \bibfield  {author} {\bibinfo {author} {\bibfnamefont {T.}~\bibnamefont
  {Oka}}\ and\ \bibinfo {author} {\bibfnamefont {H.}~\bibnamefont {Aoki}},\
  }\bibfield  {title} {\enquote {\bibinfo {title} {Photovoltaic berry curvature
  in the honeycomb lattice},}\ }\href@noop {} {\bibfield  {journal} {\bibinfo
  {journal} {Journal of Physics: Conference Series}\ }\textbf {\bibinfo
  {volume} {200}},\ \bibinfo {pages} {062017} (\bibinfo {year}
  {2010})}\BibitemShut {NoStop}%
\bibitem [{\citenamefont {Kitagawa}\ \emph {et~al.}(2010)\citenamefont
  {Kitagawa}, \citenamefont {Berg}, \citenamefont {Rudner},\ and\ \citenamefont
  {Demler}}]{Kitagawa2010}%
  \BibitemOpen
  \bibfield  {author} {\bibinfo {author} {\bibfnamefont {T.}~\bibnamefont
  {Kitagawa}}, \bibinfo {author} {\bibfnamefont {E.}~\bibnamefont {Berg}},
  \bibinfo {author} {\bibfnamefont {M.}~\bibnamefont {Rudner}}, \ and\ \bibinfo
  {author} {\bibfnamefont {E.}~\bibnamefont {Demler}},\ }\bibfield  {title}
  {\enquote {\bibinfo {title} {Topological characterization of periodically
  driven quantum systems},}\ }\href@noop {} {\bibfield  {journal} {\bibinfo
  {journal} {Phys. Rev. B}\ }\textbf {\bibinfo {volume} {82}},\ \bibinfo
  {pages} {235114} (\bibinfo {year} {2010})}\BibitemShut {NoStop}%
\bibitem [{\citenamefont {Lindner}\ \emph {et~al.}(2011)\citenamefont
  {Lindner}, \citenamefont {Refael},\ and\ \citenamefont
  {Galitski}}]{Lindner2011}%
  \BibitemOpen
  \bibfield  {author} {\bibinfo {author} {\bibfnamefont {N.~H.}\ \bibnamefont
  {Lindner}}, \bibinfo {author} {\bibfnamefont {G.}~\bibnamefont {Refael}}, \
  and\ \bibinfo {author} {\bibfnamefont {V.}~\bibnamefont {Galitski}},\
  }\bibfield  {title} {\enquote {\bibinfo {title} {{F}loquet topological
  insulator in semiconductor quantum wells},}\ }\href@noop {} {\bibfield
  {journal} {\bibinfo  {journal} {Nat. Phys.}\ }\textbf {\bibinfo {volume}
  {7}},\ \bibinfo {pages} {490} (\bibinfo {year} {2011})}\BibitemShut {NoStop}%
\bibitem [{\citenamefont {Zhou}\ and\ \citenamefont {Wu}(2011)}]{Zhou2011}%
  \BibitemOpen
  \bibfield  {author} {\bibinfo {author} {\bibfnamefont {Y.}~\bibnamefont
  {Zhou}}\ and\ \bibinfo {author} {\bibfnamefont {M.~W.}\ \bibnamefont {Wu}},\
  }\bibfield  {title} {\enquote {\bibinfo {title} {Optical response of graphene
  under intense terahertz fields},}\ }\href@noop {} {\bibfield  {journal}
  {\bibinfo  {journal} {Phys. Rev. B}\ }\textbf {\bibinfo {volume} {83}},\
  \bibinfo {pages} {245436} (\bibinfo {year} {2011})}\BibitemShut {NoStop}%
\bibitem [{\citenamefont {Kitagawa}\ \emph {et~al.}(2011)\citenamefont
  {Kitagawa}, \citenamefont {Oka}, \citenamefont {Brataas}, \citenamefont
  {Fu},\ and\ \citenamefont {Demler}}]{Kitagawa2011}%
  \BibitemOpen
  \bibfield  {author} {\bibinfo {author} {\bibfnamefont {T.}~\bibnamefont
  {Kitagawa}}, \bibinfo {author} {\bibfnamefont {T.}~\bibnamefont {Oka}},
  \bibinfo {author} {\bibfnamefont {A.}~\bibnamefont {Brataas}}, \bibinfo
  {author} {\bibfnamefont {L.}~\bibnamefont {Fu}}, \ and\ \bibinfo {author}
  {\bibfnamefont {E.}~\bibnamefont {Demler}},\ }\bibfield  {title} {\enquote
  {\bibinfo {title} {Transport properties of nonequilibrium systems under the
  application of light: Photoinduced quantum {H}all insulators without {L}andau
  levels},}\ }\href@noop {} {\bibfield  {journal} {\bibinfo  {journal} {Phys.
  Rev. B}\ }\textbf {\bibinfo {volume} {84}},\ \bibinfo {pages} {235108}
  (\bibinfo {year} {2011})}\BibitemShut {NoStop}%
\bibitem [{\citenamefont {Perez-Piskunow}\ \emph {et~al.}(2014)\citenamefont
  {Perez-Piskunow}, \citenamefont {Usaj}, \citenamefont {Balseiro},\ and\
  \citenamefont {Foa~Torres}}]{Perez-Piskunow2014}%
  \BibitemOpen
  \bibfield  {author} {\bibinfo {author} {\bibfnamefont {P.~M.}\ \bibnamefont
  {Perez-Piskunow}}, \bibinfo {author} {\bibfnamefont {G.}~\bibnamefont
  {Usaj}}, \bibinfo {author} {\bibfnamefont {C.~A.}\ \bibnamefont {Balseiro}},
  \ and\ \bibinfo {author} {\bibfnamefont {L.~E.~F.}\ \bibnamefont
  {Foa~Torres}},\ }\bibfield  {title} {\enquote {\bibinfo {title} {{F}loquet
  chiral edge states in graphene},}\ }\href {\doibase
  10.1103/PhysRevB.89.121401} {\bibfield  {journal} {\bibinfo  {journal} {Phys.
  Rev. B}\ }\textbf {\bibinfo {volume} {89}},\ \bibinfo {pages} {121401(R)}
  (\bibinfo {year} {2014})}\BibitemShut {NoStop}%
\bibitem [{\citenamefont {Usaj}\ \emph {et~al.}(2014)\citenamefont {Usaj},
  \citenamefont {Perez-Piskunow}, \citenamefont {Foa~Torres},\ and\
  \citenamefont {Balseiro}}]{Usaj2014a}%
  \BibitemOpen
  \bibfield  {author} {\bibinfo {author} {\bibfnamefont {G.}~\bibnamefont
  {Usaj}}, \bibinfo {author} {\bibfnamefont {P.~M.}\ \bibnamefont
  {Perez-Piskunow}}, \bibinfo {author} {\bibfnamefont {L.~E.~F.}\ \bibnamefont
  {Foa~Torres}}, \ and\ \bibinfo {author} {\bibfnamefont {C.~A.}\ \bibnamefont
  {Balseiro}},\ }\bibfield  {title} {\enquote {\bibinfo {title} {Irradiated
  graphene as a tunable {F}loquet topological insulator},}\ }\href {\doibase
  10.1103/PhysRevB.90.115423} {\bibfield  {journal} {\bibinfo  {journal} {Phys.
  Rev. B}\ }\textbf {\bibinfo {volume} {90}},\ \bibinfo {pages} {115423}
  (\bibinfo {year} {2014})}\BibitemShut {NoStop}%
\bibitem [{\citenamefont {Calvo}\ \emph {et~al.}(2011)\citenamefont {Calvo},
  \citenamefont {Pastawski}, \citenamefont {Roche},\ and\ \citenamefont
  {Foa~Torres}}]{Calvo2011}%
  \BibitemOpen
  \bibfield  {author} {\bibinfo {author} {\bibfnamefont {H.~L.}\ \bibnamefont
  {Calvo}}, \bibinfo {author} {\bibfnamefont {H.~M.}\ \bibnamefont
  {Pastawski}}, \bibinfo {author} {\bibfnamefont {S.}~\bibnamefont {Roche}}, \
  and\ \bibinfo {author} {\bibfnamefont {L.~E.~F.}\ \bibnamefont
  {Foa~Torres}},\ }\bibfield  {title} {\enquote {\bibinfo {title} {Tuning
  laser-induced band gaps in graphene},}\ }\href {\doibase 10.1063/1.3597412}
  {\bibfield  {journal} {\bibinfo  {journal} {Appl. Phys. Lett.}\ }\textbf
  {\bibinfo {volume} {98}},\ \bibinfo {pages} {232103} (\bibinfo {year}
  {2011})}\BibitemShut {NoStop}%
\bibitem [{\citenamefont {Wang}\ \emph {et~al.}(2013)\citenamefont {Wang},
  \citenamefont {Steinberg}, \citenamefont {Jarillo-Herrero},\ and\
  \citenamefont {Gedik}}]{Wang2013a}%
  \BibitemOpen
  \bibfield  {author} {\bibinfo {author} {\bibfnamefont {Y.~H.}\ \bibnamefont
  {Wang}}, \bibinfo {author} {\bibfnamefont {H.}~\bibnamefont {Steinberg}},
  \bibinfo {author} {\bibfnamefont {P.}~\bibnamefont {Jarillo-Herrero}}, \ and\
  \bibinfo {author} {\bibfnamefont {N.}~\bibnamefont {Gedik}},\ }\bibfield
  {title} {\enquote {\bibinfo {title} {Observation of {F}loquet-{B}loch states
  on the surface of a topological insulator},}\ }\href@noop {} {\bibfield
  {journal} {\bibinfo  {journal} {Science}\ }\textbf {\bibinfo {volume}
  {342}},\ \bibinfo {pages} {453} (\bibinfo {year} {2013})}\BibitemShut
  {NoStop}%
\bibitem [{\citenamefont {Foa~Torres}\ \emph {et~al.}(2014)\citenamefont
  {Foa~Torres}, \citenamefont {Perez-Piskunow}, \citenamefont {Balseiro},\ and\
  \citenamefont {Usaj}}]{FoaTorres2014}%
  \BibitemOpen
  \bibfield  {author} {\bibinfo {author} {\bibfnamefont {L.~E.~F.}\
  \bibnamefont {Foa~Torres}}, \bibinfo {author} {\bibfnamefont {P.~M.}\
  \bibnamefont {Perez-Piskunow}}, \bibinfo {author} {\bibfnamefont {C.~A.}\
  \bibnamefont {Balseiro}}, \ and\ \bibinfo {author} {\bibfnamefont
  {G.}~\bibnamefont {Usaj}},\ }\bibfield  {title} {\enquote {\bibinfo {title}
  {Multiterminal conductance of a {F}loquet topological insulator},}\ }\href
  {\doibase 10.1103/PhysRevLett.113.266801} {\bibfield  {journal} {\bibinfo
  {journal} {Phys. Rev. Lett.}\ }\textbf {\bibinfo {volume} {113}},\ \bibinfo
  {pages} {266801} (\bibinfo {year} {2014})}\BibitemShut {NoStop}%
\bibitem [{\citenamefont {Perez-Piskunow}\ \emph {et~al.}(2015)\citenamefont
  {Perez-Piskunow}, \citenamefont {Foa~Torres},\ and\ \citenamefont
  {Usaj}}]{Perez-Piskunow2015}%
  \BibitemOpen
  \bibfield  {author} {\bibinfo {author} {\bibfnamefont {P.~M.}\ \bibnamefont
  {Perez-Piskunow}}, \bibinfo {author} {\bibfnamefont {L.~E.~F.}\ \bibnamefont
  {Foa~Torres}}, \ and\ \bibinfo {author} {\bibfnamefont {G.}~\bibnamefont
  {Usaj}},\ }\bibfield  {title} {\enquote {\bibinfo {title} {Hierarchy of
  {F}loquet gaps and edge states for driven honeycomb lattices},}\ }\href
  {\doibase 10.1103/PhysRevA.91.043625} {\bibfield  {journal} {\bibinfo
  {journal} {Phys. Rev. A}\ }\textbf {\bibinfo {volume} {91}},\ \bibinfo
  {pages} {043625} (\bibinfo {year} {2015})}\BibitemShut {NoStop}%
\bibitem [{\citenamefont {D'Alessio}\ and\ \citenamefont
  {Rigol}(2015)}]{DAlessio2015}%
  \BibitemOpen
  \bibfield  {author} {\bibinfo {author} {\bibfnamefont {L.}~\bibnamefont
  {D'Alessio}}\ and\ \bibinfo {author} {\bibfnamefont {M.}~\bibnamefont
  {Rigol}},\ }\bibfield  {title} {\enquote {\bibinfo {title} {Dynamical
  preparation of {F}loquet {C}hern insulators},}\ }\href@noop {} {\bibfield
  {journal} {\bibinfo  {journal} {Nat Commun}\ }\textbf {\bibinfo {volume} {6}}
  (\bibinfo {year} {2015})},\ \bibinfo {note} {article}\BibitemShut {NoStop}%
\bibitem [{\citenamefont {Budich}\ and\ \citenamefont
  {Heyl}(2016)}]{Budich2016}%
  \BibitemOpen
  \bibfield  {author} {\bibinfo {author} {\bibfnamefont {J.~C.}\ \bibnamefont
  {Budich}}\ and\ \bibinfo {author} {\bibfnamefont {M.}~\bibnamefont {Heyl}},\
  }\bibfield  {title} {\enquote {\bibinfo {title} {Dynamical topological order
  parameters far from equilibrium},}\ }\href@noop {} {\bibfield  {journal}
  {\bibinfo  {journal} {Physical Review B}\ }\textbf {\bibinfo {volume} {93}}
  (\bibinfo {year} {2016})}\BibitemShut {NoStop}%
\bibitem [{\citenamefont {Sch\"{u}ler}\ and\ \citenamefont
  {Werner}(2017)}]{Schler2017}%
  \BibitemOpen
  \bibfield  {author} {\bibinfo {author} {\bibfnamefont {M.}~\bibnamefont
  {Sch\"{u}ler}}\ and\ \bibinfo {author} {\bibfnamefont {P.}~\bibnamefont
  {Werner}},\ }\bibfield  {title} {\enquote {\bibinfo {title} {Tracing the
  nonequilibrium topological state of {C}hern insulators},}\ }\href@noop {}
  {\bibfield  {journal} {\bibinfo  {journal} {Physical Review B}\ }\textbf
  {\bibinfo {volume} {96}} (\bibinfo {year} {2017})}\BibitemShut {NoStop}%
\bibitem [{\citenamefont {Dehghani}\ and\ \citenamefont
  {Mitra}(2015)}]{Dehghani2015}%
  \BibitemOpen
  \bibfield  {author} {\bibinfo {author} {\bibfnamefont {H.}~\bibnamefont
  {Dehghani}}\ and\ \bibinfo {author} {\bibfnamefont {A.}~\bibnamefont
  {Mitra}},\ }\bibfield  {title} {\enquote {\bibinfo {title} {Optical {H}all
  conductivity of a {F}loquet topological insulator},}\ }\href {\doibase
  10.1103/PhysRevB.92.165111} {\bibfield  {journal} {\bibinfo  {journal} {Phys.
  Rev. B}\ }\textbf {\bibinfo {volume} {92}},\ \bibinfo {pages} {165111}
  (\bibinfo {year} {2015})}\BibitemShut {NoStop}%
\bibitem [{\citenamefont {Dehghani}\ \emph {et~al.}(2015)\citenamefont
  {Dehghani}, \citenamefont {Oka},\ and\ \citenamefont
  {Mitra}}]{Dehghani2015b}%
  \BibitemOpen
  \bibfield  {author} {\bibinfo {author} {\bibfnamefont {H.}~\bibnamefont
  {Dehghani}}, \bibinfo {author} {\bibfnamefont {T.}~\bibnamefont {Oka}}, \
  and\ \bibinfo {author} {\bibfnamefont {A.}~\bibnamefont {Mitra}},\ }\bibfield
   {title} {\enquote {\bibinfo {title} {Out-of-equilibrium electrons and the
  {H}all conductance of a {F}loquet topological insulator},}\ }\href {\doibase
  10.1103/PhysRevB.91.155422} {\bibfield  {journal} {\bibinfo  {journal} {Phys.
  Rev. B}\ }\textbf {\bibinfo {volume} {91}},\ \bibinfo {pages} {155422}
  (\bibinfo {year} {2015})}\BibitemShut {NoStop}%
\bibitem [{\citenamefont {Wang}\ \emph {et~al.}(2016)\citenamefont {Wang},
  \citenamefont {Schmitt},\ and\ \citenamefont {Kehrein}}]{Wang2016}%
  \BibitemOpen
  \bibfield  {author} {\bibinfo {author} {\bibfnamefont {P.}~\bibnamefont
  {Wang}}, \bibinfo {author} {\bibfnamefont {M.}~\bibnamefont {Schmitt}}, \
  and\ \bibinfo {author} {\bibfnamefont {S.}~\bibnamefont {Kehrein}},\
  }\bibfield  {title} {\enquote {\bibinfo {title} {Universal nonanalytic
  behavior of the {H}all conductance in a {C}hern insulator at the
  topologically driven nonequilibrium phase transition},}\ }\href {\doibase
  10.1103/PhysRevB.93.085134} {\bibfield  {journal} {\bibinfo  {journal} {Phys.
  Rev. B}\ }\textbf {\bibinfo {volume} {93}},\ \bibinfo {pages} {085134}
  (\bibinfo {year} {2016})}\BibitemShut {NoStop}%
\bibitem [{\citenamefont {Caio}\ \emph {et~al.}(2016)\citenamefont {Caio},
  \citenamefont {Cooper},\ and\ \citenamefont {Bhaseen}}]{Caio2016}%
  \BibitemOpen
  \bibfield  {author} {\bibinfo {author} {\bibfnamefont {M.~D.}\ \bibnamefont
  {Caio}}, \bibinfo {author} {\bibfnamefont {N.~R.}\ \bibnamefont {Cooper}}, \
  and\ \bibinfo {author} {\bibfnamefont {M.~J.}\ \bibnamefont {Bhaseen}},\
  }\bibfield  {title} {\enquote {\bibinfo {title} {Hall response and edge
  current dynamics in {C}hern insulators out of equilibrium},}\ }\href@noop {}
  {\bibfield  {journal} {\bibinfo  {journal} {Physical Review B}\ }\textbf
  {\bibinfo {volume} {94}} (\bibinfo {year} {2016})}\BibitemShut {NoStop}%
\bibitem [{\citenamefont {Wilson}\ \emph {et~al.}(2016)\citenamefont {Wilson},
  \citenamefont {Song},\ and\ \citenamefont {Refael}}]{Wilson2016}%
  \BibitemOpen
  \bibfield  {author} {\bibinfo {author} {\bibfnamefont {J.~H.}\ \bibnamefont
  {Wilson}}, \bibinfo {author} {\bibfnamefont {J.~C.}\ \bibnamefont {Song}}, \
  and\ \bibinfo {author} {\bibfnamefont {G.}~\bibnamefont {Refael}},\
  }\bibfield  {title} {\enquote {\bibinfo {title} {Remnant geometric {H}all
  response in a quantum quench},}\ }\href@noop {} {\bibfield  {journal}
  {\bibinfo  {journal} {Physical Review Letters}\ }\textbf {\bibinfo {volume}
  {117}} (\bibinfo {year} {2016})}\BibitemShut {NoStop}%
\bibitem [{\citenamefont {Schmitt}\ and\ \citenamefont
  {Wang}(2017)}]{Schmitt2017}%
  \BibitemOpen
  \bibfield  {author} {\bibinfo {author} {\bibfnamefont {M.}~\bibnamefont
  {Schmitt}}\ and\ \bibinfo {author} {\bibfnamefont {P.}~\bibnamefont {Wang}},\
  }\bibfield  {title} {\enquote {\bibinfo {title} {Universal nonanalytic
  behavior of the nonequilibrium {H}all conductance in {F}loquet topological
  insulators},}\ }\href@noop {} {\bibfield  {journal} {\bibinfo  {journal}
  {Physical Review B}\ }\textbf {\bibinfo {volume} {96}} (\bibinfo {year}
  {2017})}\BibitemShut {NoStop}%
\bibitem [{\citenamefont {Hu}\ \emph {et~al.}(2016)\citenamefont {Hu},
  \citenamefont {Zoller},\ and\ \citenamefont {Budich}}]{Hu2016}%
  \BibitemOpen
  \bibfield  {author} {\bibinfo {author} {\bibfnamefont {Y.}~\bibnamefont
  {Hu}}, \bibinfo {author} {\bibfnamefont {P.}~\bibnamefont {Zoller}}, \ and\
  \bibinfo {author} {\bibfnamefont {J.~C.}\ \bibnamefont {Budich}},\ }\bibfield
   {title} {\enquote {\bibinfo {title} {Dynamical buildup of a quantized {H}all
  response from nontopological states},}\ }\href {\doibase
  10.1103/PhysRevLett.117.126803} {\bibfield  {journal} {\bibinfo  {journal}
  {Phys. Rev. Lett.}\ }\textbf {\bibinfo {volume} {117}},\ \bibinfo {pages}
  {126803} (\bibinfo {year} {2016})}\BibitemShut {NoStop}%
\bibitem [{\citenamefont {Higuchi}\ \emph {et~al.}(2017)\citenamefont
  {Higuchi}, \citenamefont {Heide}, \citenamefont {Ullmann}, \citenamefont
  {Weber},\ and\ \citenamefont {Hommelhoff}}]{Higuchi2017}%
  \BibitemOpen
  \bibfield  {author} {\bibinfo {author} {\bibfnamefont {T.}~\bibnamefont
  {Higuchi}}, \bibinfo {author} {\bibfnamefont {C.}~\bibnamefont {Heide}},
  \bibinfo {author} {\bibfnamefont {K.}~\bibnamefont {Ullmann}}, \bibinfo
  {author} {\bibfnamefont {H.~B.}\ \bibnamefont {Weber}}, \ and\ \bibinfo
  {author} {\bibfnamefont {P.}~\bibnamefont {Hommelhoff}},\ }\bibfield  {title}
  {\enquote {\bibinfo {title} {Light-field-driven currents in graphene},}\
  }\href {\doibase 10.1038/nature23900} {\bibfield  {journal} {\bibinfo
  {journal} {Nature}\ }\textbf {\bibinfo {volume} {550}},\ \bibinfo {pages}
  {224} (\bibinfo {year} {2017})}\BibitemShut {NoStop}%
\bibitem [{\citenamefont {Shirley}(1965)}]{Shirley1965}%
  \BibitemOpen
  \bibfield  {author} {\bibinfo {author} {\bibfnamefont {J.}~\bibnamefont
  {Shirley}},\ }\bibfield  {title} {\enquote {\bibinfo {title} {Solution of the
  {S}chr{\"o}dinger equation with a {H}amiltonian periodic in time},}\
  }\href@noop {} {\bibfield  {journal} {\bibinfo  {journal} {Phys. Rev.}\
  }\textbf {\bibinfo {volume} {138}},\ \bibinfo {pages} {B979} (\bibinfo {year}
  {1965})}\BibitemShut {NoStop}%
\bibitem [{\citenamefont {Sambe}(1973)}]{Sambe1973}%
  \BibitemOpen
  \bibfield  {author} {\bibinfo {author} {\bibfnamefont {H.}~\bibnamefont
  {Sambe}},\ }\bibfield  {title} {\enquote {\bibinfo {title} {Steady states and
  quasienergies of a quantum-mechanical system in an oscillating field},}\
  }\href@noop {} {\bibfield  {journal} {\bibinfo  {journal} {Phys. Rev. A}\
  }\textbf {\bibinfo {volume} {7}},\ \bibinfo {pages} {2203} (\bibinfo {year}
  {1973})}\BibitemShut {NoStop}%
\bibitem [{\citenamefont {Grifoni}\ and\ \citenamefont
  {H\"anggi}(1998)}]{Grifoni1998a}%
  \BibitemOpen
  \bibfield  {author} {\bibinfo {author} {\bibfnamefont {M.}~\bibnamefont
  {Grifoni}}\ and\ \bibinfo {author} {\bibfnamefont {P.}~\bibnamefont
  {H\"anggi}},\ }\bibfield  {title} {\enquote {\bibinfo {title} {Driven quantum
  tunneling},}\ }\href {\doibase 10.1016/S0370-1573(98)00022-2} {\bibfield
  {journal} {\bibinfo  {journal} {Phys. Rep.}\ }\textbf {\bibinfo {volume}
  {304}},\ \bibinfo {pages} {229} (\bibinfo {year} {1998})}\BibitemShut
  {NoStop}%
\bibitem [{\citenamefont {Kohler}\ \emph {et~al.}(2005)\citenamefont {Kohler},
  \citenamefont {Lehmann},\ and\ \citenamefont {H\"anggi}}]{Kohler2005}%
  \BibitemOpen
  \bibfield  {author} {\bibinfo {author} {\bibfnamefont {S.}~\bibnamefont
  {Kohler}}, \bibinfo {author} {\bibfnamefont {J.}~\bibnamefont {Lehmann}}, \
  and\ \bibinfo {author} {\bibfnamefont {P.}~\bibnamefont {H\"anggi}},\
  }\bibfield  {title} {\enquote {\bibinfo {title} {Driven quantum transport on
  the nanoscale},}\ }\href {\doibase 10.1016/j.physrep.2004.11.002} {\bibfield
  {journal} {\bibinfo  {journal} {Phys. Rep.}\ }\textbf {\bibinfo {volume}
  {406}},\ \bibinfo {pages} {379} (\bibinfo {year} {2005})}\BibitemShut
  {NoStop}%
\bibitem [{\citenamefont {Peskin}\ and\ \citenamefont
  {Moiseyev}(1993)}]{Peskin1993}%
  \BibitemOpen
  \bibfield  {author} {\bibinfo {author} {\bibfnamefont {U.}~\bibnamefont
  {Peskin}}\ and\ \bibinfo {author} {\bibfnamefont {N.}~\bibnamefont
  {Moiseyev}},\ }\bibfield  {title} {\enquote {\bibinfo {title} {The solution
  of the time-dependent {S}chr{\"o}dinger equation by the (t,t') method:
  Theory, computational algorithm and applications},}\ }\href@noop {}
  {\bibfield  {journal} {\bibinfo  {journal} {J. Chem. Phys.}\ }\textbf
  {\bibinfo {volume} {99}},\ \bibinfo {pages} {4590} (\bibinfo {year}
  {1993})}\BibitemShut {NoStop}%
\bibitem [{\citenamefont {Study}(1905)}]{Study1905}%
  \BibitemOpen
  \bibfield  {author} {\bibinfo {author} {\bibfnamefont {E.}~\bibnamefont
  {Study}},\ }\bibfield  {title} {\enquote {\bibinfo {title} {K\"{u}rzeste wege
  im komplexen gebiet},}\ }\href@noop {} {\bibfield  {journal} {\bibinfo
  {journal} {Mathematische Annalen}\ }\textbf {\bibinfo {volume} {60}},\
  \bibinfo {pages} {321} (\bibinfo {year} {1905})}\BibitemShut {NoStop}%
\bibitem [{\citenamefont {Rycerz}\ \emph {et~al.}(2007)\citenamefont {Rycerz},
  \citenamefont {Tworzyd{\l}o},\ and\ \citenamefont {Beenakker}}]{Rycerz2007}%
  \BibitemOpen
  \bibfield  {author} {\bibinfo {author} {\bibfnamefont {A.}~\bibnamefont
  {Rycerz}}, \bibinfo {author} {\bibfnamefont {J.}~\bibnamefont
  {Tworzyd{\l}o}}, \ and\ \bibinfo {author} {\bibfnamefont {C.~W.~J.}\
  \bibnamefont {Beenakker}},\ }\bibfield  {title} {\enquote {\bibinfo {title}
  {Valley filter and valley valve in graphene},}\ }\href@noop {} {\bibfield
  {journal} {\bibinfo  {journal} {Nat. Phys.}\ }\textbf {\bibinfo {volume}
  {3}},\ \bibinfo {pages} {172} (\bibinfo {year} {2007})}\BibitemShut {NoStop}%
\bibitem [{\citenamefont {Xiao}\ \emph {et~al.}(2007)\citenamefont {Xiao},
  \citenamefont {Yao},\ and\ \citenamefont {Niu}}]{Xiao2007}%
  \BibitemOpen
  \bibfield  {author} {\bibinfo {author} {\bibfnamefont {D.}~\bibnamefont
  {Xiao}}, \bibinfo {author} {\bibfnamefont {W.}~\bibnamefont {Yao}}, \ and\
  \bibinfo {author} {\bibfnamefont {Q.}~\bibnamefont {Niu}},\ }\bibfield
  {title} {\enquote {\bibinfo {title} {Valley-contrasting physics in graphene:
  Magnetic moment and topological transport},}\ }\href {\doibase
  10.1103/PhysRevLett.99.236809} {\bibfield  {journal} {\bibinfo  {journal}
  {Phys. Rev. Lett.}\ }\textbf {\bibinfo {volume} {99}},\ \bibinfo {pages}
  {236809} (\bibinfo {year} {2007})}\BibitemShut {NoStop}%
\bibitem [{\citenamefont {Sui}\ \emph {et~al.}(2015)\citenamefont {Sui},
  \citenamefont {Chen}, \citenamefont {Ma}, \citenamefont {Shan}, \citenamefont
  {Tian}, \citenamefont {Watanabe}, \citenamefont {Taniguchi}, \citenamefont
  {Jin}, \citenamefont {Yao}, \citenamefont {Xiao},\ and\ \citenamefont
  {Zhang}}]{Sui2015}%
  \BibitemOpen
  \bibfield  {author} {\bibinfo {author} {\bibfnamefont {M.}~\bibnamefont
  {Sui}}, \bibinfo {author} {\bibfnamefont {G.}~\bibnamefont {Chen}}, \bibinfo
  {author} {\bibfnamefont {L.}~\bibnamefont {Ma}}, \bibinfo {author}
  {\bibfnamefont {W.-Y.}\ \bibnamefont {Shan}}, \bibinfo {author}
  {\bibfnamefont {D.}~\bibnamefont {Tian}}, \bibinfo {author} {\bibfnamefont
  {K.}~\bibnamefont {Watanabe}}, \bibinfo {author} {\bibfnamefont
  {T.}~\bibnamefont {Taniguchi}}, \bibinfo {author} {\bibfnamefont
  {X.}~\bibnamefont {Jin}}, \bibinfo {author} {\bibfnamefont {W.}~\bibnamefont
  {Yao}}, \bibinfo {author} {\bibfnamefont {D.}~\bibnamefont {Xiao}}, \ and\
  \bibinfo {author} {\bibfnamefont {Y.}~\bibnamefont {Zhang}},\ }\bibfield
  {title} {\enquote {\bibinfo {title} {Gate-tunable topological valley
  transport in bilayer graphene},}\ }\href
  {http://dx.doi.org/10.1038/nphys3485} {\bibfield  {journal} {\bibinfo
  {journal} {Nat Phys}\ }\textbf {\bibinfo {volume} {11}},\ \bibinfo {pages}
  {1027} (\bibinfo {year} {2015})},\ \bibinfo {note} {letter}\BibitemShut
  {NoStop}%
\bibitem [{\citenamefont {Qi}\ \emph {et~al.}(2008)\citenamefont {Qi},
  \citenamefont {Hughes},\ and\ \citenamefont {Zhang}}]{Qi2008}%
  \BibitemOpen
  \bibfield  {author} {\bibinfo {author} {\bibfnamefont {X.-L.}\ \bibnamefont
  {Qi}}, \bibinfo {author} {\bibfnamefont {T.~L.}\ \bibnamefont {Hughes}}, \
  and\ \bibinfo {author} {\bibfnamefont {S.-C.}\ \bibnamefont {Zhang}},\
  }\bibfield  {title} {\enquote {\bibinfo {title} {Topological field theory of
  time-reversal invariant insulators},}\ }\href@noop {} {\bibfield  {journal}
  {\bibinfo  {journal} {Physical Review B}\ }\textbf {\bibinfo {volume} {78}}
  (\bibinfo {year} {2008})}\BibitemShut {NoStop}%
\bibitem [{\citenamefont {Rudner}\ \emph {et~al.}(2013)\citenamefont {Rudner},
  \citenamefont {Lindner}, \citenamefont {Berg},\ and\ \citenamefont
  {Levin}}]{Rudner2013}%
  \BibitemOpen
  \bibfield  {author} {\bibinfo {author} {\bibfnamefont {M.~S.}\ \bibnamefont
  {Rudner}}, \bibinfo {author} {\bibfnamefont {N.~H.}\ \bibnamefont {Lindner}},
  \bibinfo {author} {\bibfnamefont {E.}~\bibnamefont {Berg}}, \ and\ \bibinfo
  {author} {\bibfnamefont {M.}~\bibnamefont {Levin}},\ }\bibfield  {title}
  {\enquote {\bibinfo {title} {Anomalous edge states and the bulk-edge
  correspondence for periodically-driven two dimensional systems},}\
  }\href@noop {} {\bibfield  {journal} {\bibinfo  {journal} {Phys. Rev. X}\
  }\textbf {\bibinfo {volume} {3}},\ \bibinfo {pages} {031005} (\bibinfo {year}
  {2013})}\BibitemShut {NoStop}%
\bibitem [{\citenamefont {Pfeifer}\ and\ \citenamefont
  {Levine}(1983)}]{Pfeifer1983}%
  \BibitemOpen
  \bibfield  {author} {\bibinfo {author} {\bibfnamefont {P.}~\bibnamefont
  {Pfeifer}}\ and\ \bibinfo {author} {\bibfnamefont {R.~D.}\ \bibnamefont
  {Levine}},\ }\bibfield  {title} {\enquote {\bibinfo {title} {A stationary
  formulation of time-dependent problems in quantum mechanics},}\ }\href@noop
  {} {\bibfield  {journal} {\bibinfo  {journal} {The Journal of Chemical
  Physics}\ }\textbf {\bibinfo {volume} {79}},\ \bibinfo {pages} {5512}
  (\bibinfo {year} {1983})}\BibitemShut {NoStop}%
\bibitem [{\citenamefont {Breuer}\ and\ \citenamefont
  {Holthaus}(1989)}]{Breuer1989}%
  \BibitemOpen
  \bibfield  {author} {\bibinfo {author} {\bibfnamefont {H.}~\bibnamefont
  {Breuer}}\ and\ \bibinfo {author} {\bibfnamefont {M.}~\bibnamefont
  {Holthaus}},\ }\bibfield  {title} {\enquote {\bibinfo {title} {Quantum phases
  and {L}andau-{Z}ener transitions in oscillating fields},}\ }\href@noop {}
  {\bibfield  {journal} {\bibinfo  {journal} {Physics Letters A}\ }\textbf
  {\bibinfo {volume} {140}},\ \bibinfo {pages} {507} (\bibinfo {year}
  {1989})}\BibitemShut {NoStop}%
\bibitem [{\citenamefont {Drese}\ and\ \citenamefont
  {Holthaus}(1999)}]{Drese1999}%
  \BibitemOpen
  \bibfield  {author} {\bibinfo {author} {\bibfnamefont {K.}~\bibnamefont
  {Drese}}\ and\ \bibinfo {author} {\bibfnamefont {M.}~\bibnamefont
  {Holthaus}},\ }\bibfield  {title} {\enquote {\bibinfo {title} {{F}loquet
  theory for short laser pulses},}\ }\href@noop {} {\bibfield  {journal}
  {\bibinfo  {journal} {The European Physical Journal D - Atomic, Molecular and
  Optical Physics}\ }\textbf {\bibinfo {volume} {5}},\ \bibinfo {pages} {119}
  (\bibinfo {year} {1999})}\BibitemShut {NoStop}%
\bibitem [{\citenamefont {Privitera}\ and\ \citenamefont
  {Santoro}(2016)}]{Privitera2016}%
  \BibitemOpen
  \bibfield  {author} {\bibinfo {author} {\bibfnamefont {L.}~\bibnamefont
  {Privitera}}\ and\ \bibinfo {author} {\bibfnamefont {G.~E.}\ \bibnamefont
  {Santoro}},\ }\bibfield  {title} {\enquote {\bibinfo {title} {Quantum
  annealing and nonequilibrium dynamics of floquet chern insulators},}\
  }\href@noop {} {\bibfield  {journal} {\bibinfo  {journal} {Physical Review
  B}\ }\textbf {\bibinfo {volume} {93}} (\bibinfo {year} {2016})}\BibitemShut
  {NoStop}%
\bibitem [{\citenamefont {Oka}\ and\ \citenamefont
  {Bucciantini}(2016)}]{Oka2016}%
  \BibitemOpen
  \bibfield  {author} {\bibinfo {author} {\bibfnamefont {T.}~\bibnamefont
  {Oka}}\ and\ \bibinfo {author} {\bibfnamefont {L.}~\bibnamefont
  {Bucciantini}},\ }\bibfield  {title} {\enquote {\bibinfo {title} {Heterodyne
  {H}all effect in a two-dimensional electron gas},}\ }\href@noop {} {\bibfield
   {journal} {\bibinfo  {journal} {Physical Review B}\ }\textbf {\bibinfo
  {volume} {94}} (\bibinfo {year} {2016})}\BibitemShut {NoStop}%
\bibitem [{\citenamefont {Novi{\v{c}}enko}\ \emph {et~al.}(2017)\citenamefont
  {Novi{\v{c}}enko}, \citenamefont {Anisimovas},\ and\ \citenamefont
  {Juzeli{\={u}}nas}}]{Novienko2017}%
  \BibitemOpen
  \bibfield  {author} {\bibinfo {author} {\bibfnamefont {V.}~\bibnamefont
  {Novi{\v{c}}enko}}, \bibinfo {author} {\bibfnamefont {E.}~\bibnamefont
  {Anisimovas}}, \ and\ \bibinfo {author} {\bibfnamefont {G.}~\bibnamefont
  {Juzeli{\={u}}nas}},\ }\bibfield  {title} {\enquote {\bibinfo {title}
  {Floquet analysis of a quantum system with modulated periodic driving},}\
  }\href@noop {} {\bibfield  {journal} {\bibinfo  {journal} {Physical Review
  A}\ }\textbf {\bibinfo {volume} {95}} (\bibinfo {year} {2017})}\BibitemShut
  {NoStop}%
\bibitem [{\citenamefont {Obraztsov}\ \emph {et~al.}(2014)\citenamefont
  {Obraztsov}, \citenamefont {Kaplas}, \citenamefont {Garnov}, \citenamefont
  {Kuwata-Gonokami}, \citenamefont {Obraztsov},\ and\ \citenamefont
  {Svirko}}]{Obraztsov2014}%
  \BibitemOpen
  \bibfield  {author} {\bibinfo {author} {\bibfnamefont {P.~A.}\ \bibnamefont
  {Obraztsov}}, \bibinfo {author} {\bibfnamefont {T.}~\bibnamefont {Kaplas}},
  \bibinfo {author} {\bibfnamefont {S.~V.}\ \bibnamefont {Garnov}}, \bibinfo
  {author} {\bibfnamefont {M.}~\bibnamefont {Kuwata-Gonokami}}, \bibinfo
  {author} {\bibfnamefont {A.~N.}\ \bibnamefont {Obraztsov}}, \ and\ \bibinfo
  {author} {\bibfnamefont {Y.~P.}\ \bibnamefont {Svirko}},\ }\bibfield  {title}
  {\enquote {\bibinfo {title} {All-optical control of ultrafast photocurrents
  in unbiased graphene},}\ }\href@noop {} {\bibfield  {journal} {\bibinfo
  {journal} {Scientific Reports}\ }\textbf {\bibinfo {volume} {4}} (\bibinfo
  {year} {2014})}\BibitemShut {NoStop}%
\end{thebibliography}

\end{document}